# Quantum Modeling of Filter Bubbles Based on Kubo-Matsubara Form Green's Functions Considering Remote and Proximity Interactions:Ultraviolet Divergence to Indefinite Ghosting, Consideration of Cut Surfaces

Yasuko Kawahata [†]

Faculty of Sociology, Department of Media Sociology, Rikkyo University, 3-34-1 Nishi-Ikebukuro,Toshima-ku, Tokyo, 171-8501, JAPAN.
ykawahata@rikkyo.ac.jp

**Abstract:** The purpose of this study is to analyze the generation of filter bubbles and their social consequences observed in digital and offline environments using mathematical methods of field quantum theory. The model applies the principles of remote interaction and proximity interaction to track opinion evolution and population dynamics among agents. In particular, we incorporate non-physical factors including false alarms and confirmation bias as FP ghosting phenomena in order to mathematically represent opinion resonance and echo chamber effects within filter bubbles. Furthermore, we use the uncertainty ghosting phenomenon, a social science concept similar to the uncertainty principle, to model information uncertainty and nonlinearities in opinion formation.This research also introduces the Matsubara form of the Kubo equation and the Green function to mathematically represent temporal effects. This allows us to model how past, present, and future opinions interact and to identify the mechanisms of opinion divergence and aggregation. We simulate the formation and growth of filter bubbles and their progression to ultraviolet divergence phenomena using multiple parameters, including population density and the extreme values of opinions generated on a random number basis. By observing opinion resonance and disconnection within a society via the disconnection function, we introduce our hypothesis and discussion on regional differences in media coverage and its effectiveness in Japan, a disaster-prone country.Furthermore, we use critical points on the spin glass in simulations of filter bubble outbreaks and discuss scenarios with remote and close interactions from a social science perspective. In particular, we focus on the ultraviolet divergence and FP ghosting perspectives, discussing their advantages and disadvantages in the event of a disaster, as well as sensitive video and image considerations. This approach addresses the challenging task of applying quantum field theory to the social sciences and provides new insights into social phenomena. However, interpretation of the results requires careful consideration, and empirical validation is an issue for the future.

## 1. Introduction

Analyzing the social context and challenges of filter bubbles within the framework of quantum field theory is an interdisciplinary research area at the intersection of sociology and physics. This research attempts to mathematically describe the processes of filter bubble formation, maintenance, and destruction through quantum theoretical modeling of social interactions. The lack of media literacy and information immunity faced by digital natives is seen as the main driver of filter bubble formation, which may pave the way for social

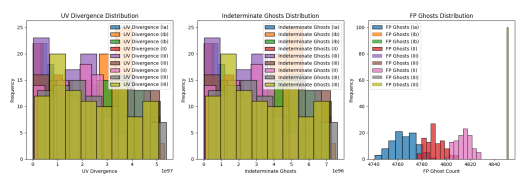

Fig. 1: Ex.Village FP Ghosts, UV Divergence, Indeterminate Ghosts score (per opinion divergence / disconnect scenario)

fragmentation and exploitation.

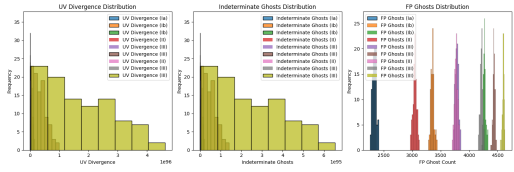

Fig. 2: Ex.City FP Ghosts, UV Divergence, Indeterminate Ghosts score (per opinion divergence / disconnect scenario)

Quantum field theory's concepts of remote interaction and proximity interaction are applied to explain two distinct social contexts: interaction in digital environments, where information propagates across physical distances, and interaction in face-to-face relationships. These theoretical approaches provide a framework for elucidating the dynamics of social communication and understanding patterns of information propagation and the process of opinion formation.

In mathematical modeling of filter bubbles, remote interactions on the propagation of information and opinions are well suited to represent global communication via social media and the Internet. In contrast, proximity interactions represent the propagation of information in a community or face-to-face setting. Such interactions are essential to understanding the subtle dynamics of social communication.

## 1.1 FP ghosting and indeterminate ghosting phenomena

FP ghosting and indeterminate ghosting phenomena are considered when modeling non-physical elements such as misinformation and confirmation bias in social communication. These phenomena are recognized as factors that can cause echo chamber effects and resonance of opinions within the filter bubble.

The development of mathematical models based on quantum field theory offers a new perspective on the social sciences and has the potential to lead to a deeper understanding of the filter bubble phenomenon. The model uses Feynman diagrams and loop diagrams to explain opinion circulation and self-referential feedback loops, presenting a new approach to understanding information propagation and opinion formation in social phenomena.

It aims to explore the mechanisms of filter bubbles in digital and offline environments and analyze their social consequences. This research crosses the boundaries between the social sciences and the physical sciences It is an attempt to cross over and open new avenues for a better understanding of complex social phenomena. It is a step toward exploring the quantum aspects of social interaction and mathematically representing the dynamics of social communication and information propagation.

The social phenomenon of filter bubbles, which can be found not only in online spaces but also in offline communities, refers to the process by which certain information and opinions are echoed and reinforced within a closed group. This phenomenon can create an echo chamber effect of information and promote social fragmentation. As a new approach to this sociological problem, this study attempts to mathematically model filter bubbles by applying the concepts of quantum field theory. This approach aims at a deeper understanding of social phenomena through the quantum mechanical interplay between information propagation and opinion formation.

## 1.2 Remote and proximity interactions

Remote and proximity interactions are key concepts in quantum field theory and are used in physics to describe the action of forces between particles. In this study, we apply these concepts to the context of the social sciences to model how information and opinions propagate within social networks and influence the formation of collective opinions. Distance interaction represents the exchange of information across non-physical and physical distances due to the prevalence of social media and digital communication. Proximity interaction, on the other hand, refers to direct communication between people based on geographic proximity.

FP ghosting and indefinite ghosting phenomena are emerging concepts that model misinformation, bias, and uncertainty in social discussions. These phenomena are important to society's understanding of information quality and reliability and provide a framework for exploring the internal dynamics of filter bubbles.

The innovation of this research lies in the application of quantum field theory, based on the concepts of remote interaction and proximity interaction, to develop a mathematical modeling and predictive framework for social phenomena. This approach aims to add a new dimension to current social science methods.

The mathematical model of the filter bubble incorporates a probabilistic approach to social interaction and is designed to capture the dynamics of opinion formation and change. The model captures the process by which individual agents exchange opinions through their interactions, generating a collective flow of opinions.

Finally, we will share our hypotheses and considerations for the model case of this paper, which is a close examination of regional differences in media coverage and its effectiveness and considerations unique to Japan, a disaster-prone country.

## 1.3 The Matsubara form and The Kubo formula, and applying the causal Green's function

In this paper, in addition, we analyze the temporal evolution of opinions in the filter bubble phenomenon by utilizing the Matsubara form and the Kubo formula, and applying the causal Green's function, the delayed Green's function, and the advanced Green's function in a social science context. By doing so, we seek to understand how information circulates and opinions are formed through digital and analog societies. Modeling influences across time and space is an attempt to delve deeper into the formation of filter bubbles and their effects.

The causal Green function deals with the impact of past events on the present. This function shows how an opinion at a given point in time interacts with past opinions. The delayed Green's function models the impact of current events on the future, while the advanced Green's function shows the inverse impact of future events on the past. In the Matsubara form, path integrals are used to treat the dynamics of quantum systems in full time, and the Kubo formula describes the response function of a physical quantity in thermodynamic equilibrium.

The application of these mathematical concepts in the social sciences allows us to quantify the complex processes of opinion formation and propagation. For example, causal Green functions can be used to analyze how past trends and events influence current public opinion. A lagged Green's function can also help predict how current policy decisions and social trends will affect society in the future.

As a modeling of spatio-temporal influences, the introduction of field quantum theory into the filter bubble discussion is useful for capturing the complex interplay between the temporal evolution and spatial diffusion of opinions. In a digital society, information can spread instantly around the world, whereas in an analog society, the propagation of opinions is limited by time and space. Replicating these differences in a model requires a deep understanding of the mechanisms of information propagation and opinion formation. Models that apply quantum field theory provide a powerful framework for mathematically representing these dynamics. Using the tools of quantum field theory, including the Matsubara form and the Kubo formula, we model the temporal change and spatial propagation of opinions in digital and analog societies. We propose to open up a new research area in the social sciences by This suggests the possibility of gaining new insights by analyzing social science problems from a physical science perspective.

This paper is submitted as a discussion paper, the purpose of which is to stimulate discussion and provide new insights at the intersection of social and physical sciences. The proposed model provides a framework for understanding the growth of filter bubbles and their social consequences through the phenomenon of ultraviolet divergence, but it is only a theoretical model and not a real social phenomenon. What this research provides is a new perspective to view social phenomena through the lens of quantum field theory, and further research is needed to apply it to actual social issues.

# 2. Ultraviolet Divergence and Ultraviolet Completion in Social Science Models

Ultraviolet divergence in the context of the social sciences is a valuable concept, serving as a metaphor, even though it differs from the physical ultraviolet divergence encountered in quantum field theory. In quantum field theory, ultraviolet divergence refers to the issue where perturbation theory calculations lead to infinite results, signifying a fundamental inconsistency within the theory. To address this problem, a technical process called regularization and renormalization is employed, which modifies the theory to ensure that observable physical quantities remain meaningful. This process is referred to as "ultraviolet completion."

In social science models, the term "ultraviolet divergence" may be used to describe model behavior that exhibits unrealistic or irrational outcomes in opinion formation, information propagation, and social network dynamics. This can occur in extreme cases, such as the rapid spread of information or the instant formation of filter bubbles. It may indicate that social science theories produce results that lack validity at small scales, such as the individual or small group level.

## 2.1 Reading "Ultraviolet Completion" in Social Science Models

**Regularization**: In social models, "regularization" involves adjusting model parameters to constrain the model's behavior within reasonable bounds. For instance, imposing upper limits on the strength or rate of opinion propagation prevents the model from exhibiting extreme behavior.

**Renormalization**: "Renormalization" in social models refers to the process of readjusting the model to mitigate extreme behavior. This may involve modifying the strength of social influences to prevent the emergence of extreme opinions or introducing diverse sources of information to avoid one-sided information flows.

**Why Local Field Theory is Used**: Local field theory is employed in the social sciences because it aligns well with observed data. For example, if a theory of opinion propagation or social influence can successfully explain an actual social phenomenon, it is considered a valuable and useful theory.

The presence of "ultraviolet divergence" problems in social science models may suggest that a theory is inadequate in explaining real social phenomena. This can occur when the model is overly simplistic or overlooks certain aspects of social interaction. By modifying the model and achieving "ultraviolet completion," more realistic social science theories can be developed.

# 3. Discussion:UV Divergence

## 3.1 UV Divergence in Physics and Social Science Models

UV divergence is a phenomenon in physics where the theory becomes infinite or uncontrollable in the high energy (short range) region. This is usually seen in perturbation calculations in quantum field theories, where the fields interact strongly on very small scales.

While the term "ultraviolet divergence" may be used metaphorically in social science models, it is possible to see an analogue of "ultraviolet divergence" in simulations of opinion propagation. This may refer to situations where information and opinions are extremely amplified within a very localized and small community or network, creating a filter bubble or echo chamber phenomenon.

For example, on social media platforms, misinformation can spread rapidly within a small group, and the intensity of individual opinions within that group can become very strong. Such a phenomenon can lead to increased social polarization and can be a potential threat to social harmony.

> **Rapid Change in Agents' Opinions** The phenomenon of some agents in a simulation rapidly leaning toward extreme opinions. This suggests that they are affected by a strong FP ghost term, akin to "ultraviolet divergence" in social models.
>
> **Applying a Type Ia Disconnection** In this scenario, the exchange of opinions is strongly restricted, leading to the formation of local echo chambers.
>
> **Applying Type Ib Disconnection** Disconnection with York decomposition applied, in which the change of opinion is restricted only under certain conditions, may lead to localized phenomena in which certain agents or groups strongly believe misinformation.

These examples illustrate how complex the propagation of information is in social networks and how it can be greatly affected by local dynamics. Understanding phenomena such as "ultraviolet divergence" in simulation models is important for understanding how misinformation and extreme opinions affect society and for developing policies and strategies to deal with them.

## 3.2 Ultraviolet Divergence in Social Science Models

Ultraviolet divergence (UV divergence) in quantum field theory refers to the problem of physical quantities diverging to infinity in high-energy (short-range) behavior. This problem usually appears in particle physics theories such as quantum electrodynamics (QED) and quantum chromodynamics (QCD), but similar concepts can be applied to simulations in the social sciences. Various scenarios in which ultraviolet divergence occurs in social science models are listed below:

> **Reinforcement of Opinion** The phenomenon in which information shared among agents is reinforced by the echo chamber effect, resulting in a sharp polarization of opinion within a small group. This can lead to deviations from social consensus as certain beliefs and prejudices become very strong.
>
> **Information Clustering** The phenomenon in which agents holding the same opinion are closely clustered within a social network, eliminating differing opinions. In this scenario, certain information may be confined to a few groups, preventing its diffusion throughout society.
>
> **Information Cascade** The phenomenon in which the opinions of some agents, through their disproportionate influence, spread quickly to other agents, causing large-scale social change. This process can lead to changes in social opinion through rapid diffusion, regardless of whether the information is true or not.
>
> **Echo Chambers and Information Disconnection** The phenomenon in which a particular community becomes disconnected from external sources of information, reinforcing only internal information, resulting in a loss of diversity throughout society. In this scenario, the filter bubble is reinforced, and a social disconnection of opinion may develop.
>
> **Opinion Interaction and Selective Exposure** The phenomenon in which opinions become polarized within a social network as agents selectively accept only information that is consistent with their beliefs. This scenario can lead to a decrease in social opinion diversity and an increase in extreme opinions.

# 4. Discussion:Applying the Replica Method

## 4.1 Ultraviolet Divergence in Social Science Models

In social science contexts, simulations that mimic ultraviolet divergence may use various cut-off functions to control for sudden local changes in opinion formation and information propagation. These cut-off functions are used to model how to incorporate the effects of non-physical factors and misinformation in opinion exchange.

## 4.2 Ultraviolet Divergence for Type Ia Disconnections (Strong Constraints)

This type of disconnection places very strict constraints on the exchange of ideas. If ultraviolet divergence were to occur in a simulation, it might manifest itself as a localized phenomenon in which certain information or beliefs are extremely reinforced in an environment where opinion change is very restricted. For example, this might be a situation where information circulates only within a particular community, and outside information is completely blocked out.

## 4.3 Ultraviolet Divergence in the Case of Type Ib Disconnection (York Decomposition)

In the case of Ib-type disconnection, the analysis is performed using the York decomposition. The York decomposition would provide a mechanism to ensure that certain beliefs and information are strongly held by the agents while preserving some of the interaction between agents.

## 4.4 Ultraviolet Divergence for Type II Disconnections (Mild Constraints)

In a Type II disconnection, which represents a mild constraint, the exchange of opinions occurs more freely, but misinformation and extreme opinions may spread within the social network to some extent. Ultraviolet divergence can occur in this scenario as a phenomenon in which misinformation spreads rapidly among certain agents or groups.

## 4.5 Ultraviolet Divergence in the Case of Type III Disconnection (Active Exchange Under Certain Conditions)

A Type III disconnection indicates that opinions are actively exchanged only under certain conditions. Ultraviolet divergence in this disconnection may manifest itself as a rapid spread of information and opinions on only certain topics and beliefs, while other topics and beliefs are ignored.

Ultraviolet divergence in these scenarios actually represents localized nonlinear dynamics and abrupt changes in the opinion formation process, providing a mathematical approach to understanding real-world phenomena such as filter bubble and echo chamber formation and opinion extremes. Simulations based on these truncation functions show how misinformation and extremes of opinions are useful in understanding how they are reinforced within individual agents and communities and in developing countermeasures against them.

To propose a way of thinking in social science models when ultraviolet divergence occurs, it is important to use the concept of ultraviolet divergence in physics as a metaphor to try to understand local extreme behavior in social dynamics. The following are suggested ideas in the application of ultraviolet divergence to the social sciences.

## 4.6 Suggested Ideas for Applying Ultraviolet Divergence in Social Sciences

(1) **Local Extremes of Perception** If ultraviolet divergence occurs in models of opinion formation and information dissemination, it can represent a situation in which extreme opinions and misinformation are rapidly amplified within a particular community or small group. In groups with short social distance (strong proximity interaction), certain misinformation may be extremely reinforced, forming echo chambers.

(2) **Faster Information Propagation and Oversaturation** In remote interactions, platforms such as social media can spread information quickly, which can lead to a phenomenon similar to ultraviolet divergence. Some information may spread instantly, causing information oversaturation in society as a whole.

(3) **Disconnection of Opinion Exchange and Filter Bubbles** When strong constraints are imposed, such as Type Ia and Type III disconnections, social interaction is restricted, and the exchange of ideas is blocked. This can create a filter bubble and cause internal amplification of misinformation. Ultraviolet divergence can represent local extremes of opinion under these constraints.

(4) **Simulation and Model Modification** Assuming that ultraviolet divergence does occur, it is necessary to study this phenomenon through simulation and to modify the model. Using the replica method, analyze the statistical properties across different replicas to determine which factors are causing the UV divergence.

(5) **Mitigation Measures for Ultraviolet Divergence** To address the social problems caused by ultraviolet divergence, specific measures should be taken to improve information literacy, ensure transparency, and promote access to diverse sources of information. This will alleviate extreme polarization in social opinion formation and promote a healthier information environment.

Applying the concept of ultraviolet divergence to the social sciences can provide a new perspective for understanding local nonlinear dynamics in opinion formation and information dissemination and for taking concrete measures to address them.

The ideas of remote and close interactions in quantum field theory and the application of the replica method to models in the social sciences where ultraviolet divergence occurs and disconnections are critical for understanding and mitigating extreme behaviors and misinformation in society.

## 4.7 Application to Types of Disconnections

### 4.7.1 For Type Ia Disconnections (Strong Constraints)

When ultraviolet divergence occurs and Type Ia disconnection is applied, it represents a situation where very strong opinions or beliefs are rapidly amplified within some agents or groups. This disconnection restricts the selective propagation of information and opinions within a social network, and certain misinformation and extreme beliefs may be locally reinforced, creating polarization.

#### 4.7.2 In the Case of Type Ib Disconnection (York Decomposition)

Type Ib truncation provides finer control over the behavior of the system through the analysis of UV fixed points. When ultraviolet divergence occurs, this approach can reveal the structure of the interaction such that misinformation only has an effect among certain agents, while others are less affected.

#### 4.7.3 In the Case of Type II Disconnection (Mild Constraint)

Type II disconnection places relatively mild constraints on the exchange of opinions, but when ultraviolet divergence occurs, information and opinions are more likely to propagate over a wide area. In this case, it is important to avoid localized concentration of misinformation by increasing interaction between different agents to mitigate the effects of ultraviolet divergence.

#### 4.7.4 In the Case of Type III Disconnection (Active Exchange Under Specific Conditions)

Type III disconnection is a setting in which the exchange of opinions is active only under certain conditions. Using this disconnection, it is possible to limit the impact of ultraviolet divergence to a localized area, thereby limiting its impact on society as a whole. This is particularly useful when modeling phenomena where misinformation and extreme opinions are only reinforced in relation to specific topics or events.

#### 4.7.5 Applying the Replica Method

Using the replica method, the statistical properties of a scenario in which ultraviolet divergence occurs can be analyzed through multiple replicas to average the effects of ultraviolet divergence in different disconnected scenarios. This allows us to understand how model parameters and forms of interaction affect ultraviolet divergence and to gain insight into designing more effective interventions for information dissemination and opinion formation. These ideas will help us to better understand the nonlinear dynamics and extreme behavior in models of information propagation and opinion formation in the social sciences and provide a basis for developing practical strategies to address the problems of misinformation and filter bubbles.

## 5. Discussion:Ultraviolet Divergence and FP Ghost Terms in Social Science Models

To comprehend the computational processes in social science models involving ultraviolet divergence and FP ghost terms, it is essential to translate these concepts into a social science context and analyze their specific behavior in different scenarios. The following scenarios and computational processes for ultraviolet divergence and FP ghost terms in social science models are considered:

### 5.1 Ultraviolet Divergence Analogy and Computational Process

**Interpretation of Ultraviolet Divergence**: Ultraviolet divergence is interpreted in social science models as a phenomenon signifying sudden changes or extremes of opinion. It implies that an agent's opinion can be infinitely amplified under certain conditions or influences.

**Computational Process for Ultraviolet Divergence**: In modeling interactions between agents, the computational process is designed in a way that if the reinforcement of an opinion surpasses a certain threshold, the opinion undergoes uncontrollable amplification. This corresponds to some agents being strongly influenced by extreme opinions or misinformation, such as receiving information solely from specific sources while ignoring others.

### 5.2 The Impact of the FP Ghost Term and the Calculation Process

**Modeling the FP Ghost Term**: The FP ghost term models non-physical elements like misinformation and prejudice in opinion propagation among agents.

**Computational Process for FP Ghost Terms**: The computational process involves updating opinions by introducing the influence of ghost terms during interactions between agents. This process accounts for the possibility that a particular agent may be influenced by misinformation or prejudice, resulting in a change of opinion. The strength of the ghost term represents the intensity of the influence of these non-physical factors and is adjusted during the simulation.

### 5.3 Application of the Cutting Function and Computational Process

**Modeling the Cutting Function**: The cutting function represents constraints on the exchange of ideas, meaning that interactions between agents are allowed only under certain conditions.

**Computational Process for Cutting Functions**: For example, in the case of a Type Ia disconnection, the computational process strictly restricts the exchange of ideas, increasing the likelihood of forming local echo chambers. The formula may be expressed as:

$$O_i^{(new)} = O_i + \delta \cdot (I_i + G_{ij}) \cdot R_k^{Ia}(O_i, O_j)$$

Where $\delta$ is the update step size, $I_i$ represents the opinion intensity, $G_{ij}$ denotes the ghost term influence, and $R_k^{Ia}(O_i, O_j)$ is the cutoff function that takes values of 1 under specific conditions, otherwise 0.

By incorporating these scenarios and computational processes into social science models, a more comprehensive understanding of information propagation dynamics, opinion formation, and social interactions can be achieved. Furthermore, this can provide fresh insights into addressing issues like misinformation and filter bubbles.

To comprehend the computational process of ultraviolet divergence and FP ghost terms in social science models, it is crucial to express these concepts mathematically and analyze model behavior accordingly. The following provides a description of the computational process based on these concepts with mathematical formulas.

# 6. Calculation of Ultraviolet Divergence and FP Ghost Terms using Entanglement Entropy, Spin States, and Relative Entropy

In this section, we will explore the calculation of ultraviolet divergence and FP ghost terms in a social science model using concepts from quantum physics, specifically entanglement entropy, spin states, and relative entropy.

## 6.1 Introduction to Entanglement Entropy

### 6.1.1 Mathematical Definition

Entanglement entropy ($S_{\text{ent}}$) serves as a measure of the complexity of information exchange between agents. It is calculated based on the degree of quantum entanglement in interactions between agents. The mathematical definition is as follows:

$$S_{\text{ent}} = -\sum_i p_i \log p_i$$

Here, $p_i$ represents the probability distribution of the opinion of agent $i$.

### 6.1.2 Computation Process

During the simulation, the probability distribution of each agent's opinion ($p_i$) is continuously updated. Entanglement entropy ($S_{\text{ent}}$) is calculated based on these probability distributions. This quantitative measure allows us to evaluate the complexity of information exchange between agents.

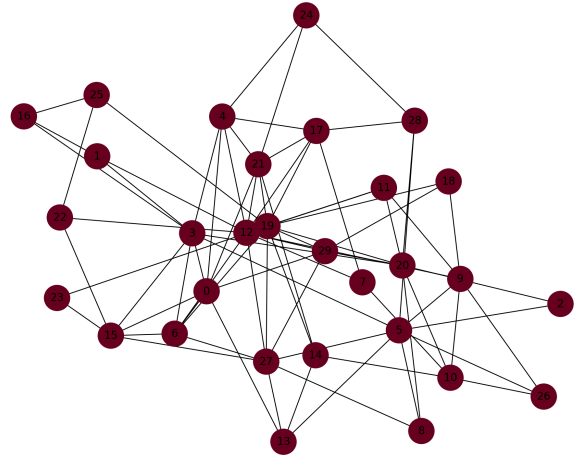

Fig. 3: Network Graph after Opinion Updates

## 6.2 Introducing the State of Spin

### 6.2.1 Definition of the Formula

The state of spin ($S_i$) is employed as a measure of the "orientation" of agents' opinions. It is determined by the sign of each agent's opinion ($O_i$), represented as follows:

$$S_i = \text{sign}(O_i)$$

### 6.2.2 Computation Process

As agents' opinions ($O_i$) are updated during the simulation, the corresponding spin state ($S_i$) is also updated. This allows us to track whether agents' opinions are trending positively or negatively.

## 6.3 Introducing Relative Entropy

### 6.3.1 Definition of the Formula

Relative entropy ($D_{\text{KL}}(P||Q)$) is utilized as a measure of the difference between two probability distributions ($P$ and $Q$). The formula for relative entropy is as follows:

$$D_{\text{KL}}(P||Q) = \sum_i P(i) \log \frac{P(i)}{Q(i)}$$

Here, $P(i)$ and $Q(i)$ represent the probability distributions of opinions within different agent groups.

During the simulation, we calculate the probability distributions of opinions for different agent groups ($P(i)$ and $Q(i)$). Using these distributions, we compute the relative entropy ($D_{\text{KL}}(P||Q)$) to quantitatively assess the disparities in opinions among various groups.

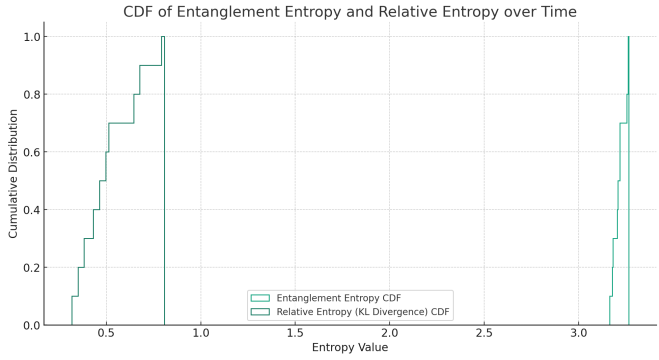

Fig. 4: CDF of Entanglement Entropy and Relative Entropy over Time

### Entanglement Entropy ($S_{ent}$) : 3.298

This is a measure of the complexity of information exchange between agents. This value indicates the degree to which the distribution of opinions among agents is complex and entangled.

### Relative Entropy (KL Divergence) ($D_{KL}$) : 0.351

This is a measure of the difference in the distribution of opinions between two different groups of agents. A larger value indicates a larger difference in the distribution of opinions between the groups.

The network graph shows the state of opinion among agents after a time step. The colors of the nodes in the graph (ranging from red to blue) indicate the value of each agent's opinion, reflecting the result of the opinion being updated.

This simulation provides insight into the dynamics of opinion formation and information exchange among agents. The values of entanglement entropy and relative entropy can be used to quantitatively assess the complexity of information exchange in the model and the differences in opinions between groups.

### 6.4 Results Description of the Social Science Model

In this section, we will describe the results obtained from the social science model, focusing on how entanglement entropy and relative entropy (Kullback–Leibler divergence) are modeled and how they change over time.

### 6.5 Cumulative Distribution Function (CDF) Graph

The CDF graph provides insights into the distribution of entanglement entropy and relative entropy values over time. Discrete jumps in entropy values at different time steps are observed in the CDF. The entanglement entropy CDF indicates that, at certain points, the complexity of information exchange between agents experiences sudden increases. Similarly, the relative entropy CDF demonstrates significant changes in the difference between the two probability distributions, P and Q, at specific time points, indicating shifts in opinion distribution among different agent groups.

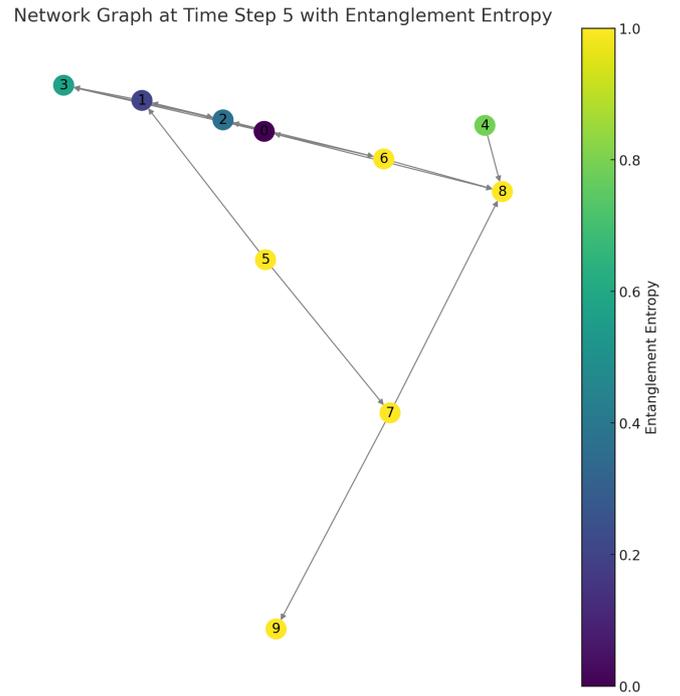

Fig. 5: Network Graph at Time Step with Entanglement Entropy

### 6.6 Network Graph

The network graph illustrates entanglement entropy among agents at a specific time step, for example, Time Step 5. Nodes are colored based on their entanglement entropy, suggesting that certain nodes (agents) have more complex interactions than others. The variation in colors can indicate clusters of agents that are tightly connected or exhibit more complex opinion dynamics.

### 6.7 Interpretation in the Context of Social Science Models

In the context of social science models, entanglement entropy quantifies the complexity of opinion dynamics within a network. Higher entanglement entropy values may signify communities where opinions are highly intertwined, potentially making them more susceptible to rapid shifts due to new information or influential agents.

Relative entropy measures the differences in opinion distributions between two subgroups within the network. Significant divergence suggests polarization or the existence of echo chambers, where one group's opinions do not overlap with the other, leading to a lack of common ground.

### 6.8 FP Ghost Term and Entanglement Entropy

The FP ghost term, represented by $G_{ij}$, influences opinion dynamics by introducing non-physical elements like misinformation or bias. When considering the FP ghost term's effect on entanglement entropy in the network graph, we may observe that certain misinformation campaigns or biases can lead to increased complexity in information sharing within the network. This can result in specific nodes (agents) having disproportionately high levels of entanglement entropy compared to others.

### 6.9 Considerations for Simulation and Model Analysis

When simulating such models, the initial conditions and parameter values, including the adaptation rate ($\delta$), interaction strength ($\beta_{ij}$), and ghost term strength ($\gamma$), are crucial as they significantly influence dynamics and resulting entropies. Random assignment of these values can serve as a starting point for hypothesis testing, but precise calibration is necessary for the model to offer meaningful insights into real-world social dynamics.

Incorporating these concepts into a computational model provides a robust framework for understanding and analyzing complex social phenomena such as the spread of misinformation, opinion polarization, and the dynamics of social consensus. By visualizing these processes through graphs and monitoring changes over time, researchers can identify patterns, test interventions, and explore the conditions under which social harmony or discord may emerge.

### Analysis of Heat Maps in a Social Science Model

Results appear to be heat maps representing the evolution of certain quantities over time in a social science model. The first image seems to be a composite heat map of "Entanglement Entropy and FP Ghost Term Over Time," while the second one is split into two separate heat maps, "Opinion Evolution Over Time" on top and "FP Ghost Terms Evolution Over Time" at the bottom. Here's a detailed analysis of each component:

### Entanglement Entropy

1. **Definition**: Entanglement entropy $S_{\text{ent}}$ measures the complexity of information exchange between agents. It is defined as:

$$S_{\text{ent}} = -\sum_i p_i \log p_i$$

where $p_i$ represents the probability distribution of agent $i$'s opinion.

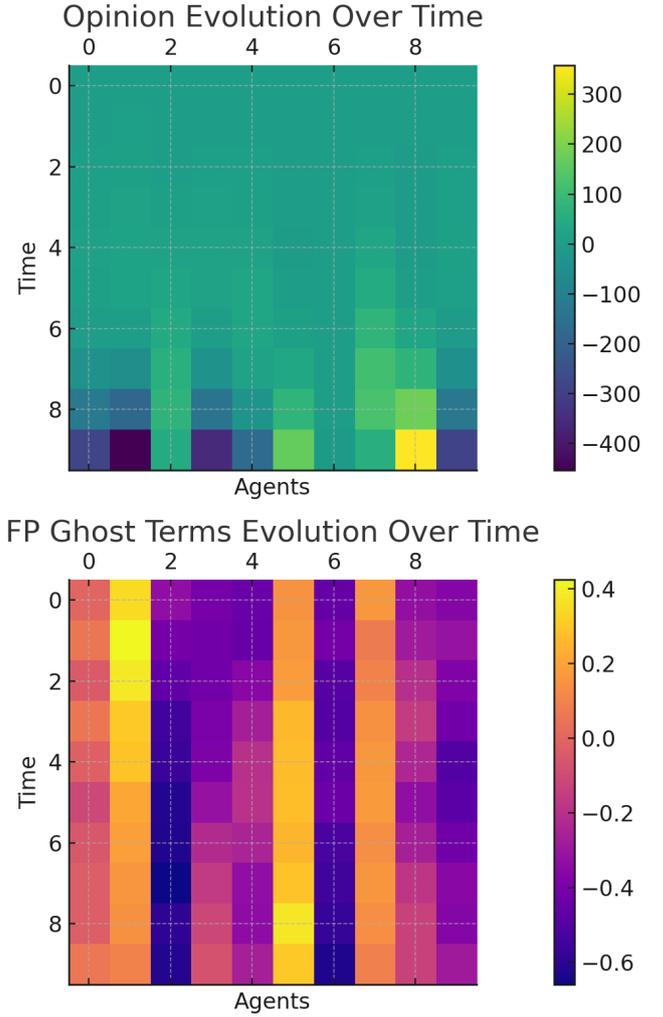

Fig. 6: Opinion$_F PGhostTermsEvolutionOverTime$

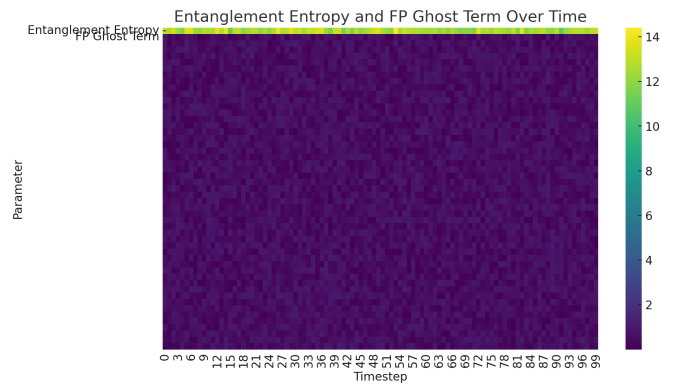

Fig. 7: Entanglement Entropy and FP Ghost Term Over Time

2. **Calculation**: During simulations, each agent's opinion probability distribution $p_i$ is updated, and the entanglement entropy $S_{\text{ent}}$ is calculated based on this. This quantitatively assesses the complexity of information exchange between agents.

**Spin States**

1. **Definition**: The spin state $S_i$ indicates the 'direction' of agent $i$'s opinion and is defined as:

$$S_i = \text{sign}(O_i)$$

where $\text{sign}(O_i)$ denotes the sign of agent $i$'s opinion $O_i$.

2. **Calculation**: With each update of an agent's opinion $O_i$, the spin state $S_i$ is also updated. This allows for tracking the positive or negative trends in agents' opinions.

**Relative Entropy**

1. **Definition**: The relative entropy $D_{\text{KL}}(P||Q)$ measures the difference between two probability distributions $P$ and $Q$, and is defined as:

$$D_{\text{KL}}(P||Q) = \sum_i P(i) \log \frac{P(i)}{Q(i)}$$

where $P(i)$ and $Q(i)$ represent the probability distributions of opinions from different agent groups.

**Analysis of the Heat Maps**

**First Image (Entanglement Entropy and FP Ghost Term Over Time)**: This heat map appears to show the evolution of a combined metric over time and across various parameters. The x-axis, labeled 'Timestep', likely represents the progression of time during the simulation, while the y-axis, labeled 'Parameter', could represent different agents or settings under which the simulation is run. The colors represent the magnitude of the combined entanglement entropy and FP ghost term, with lighter colors indicating higher values.

**Second Image (Opinion Evolution and FP Ghost Terms Evolution)**:

– **Top Heat Map (Opinion Evolution Over Time)**: This shows the changes in opinions over time across different agents. The x-axis, labeled 'Agents', represents different individuals or entities, and the y-axis represents time. The color scale shows the magnitude of opinions, with extreme values shown in purple and yellow, indicating a wide variance in opinion states over time.

– **Bottom Heat Map (FP Ghost Terms Evolution Over Time)**: This shows the evolution of the FP ghost term, which in some contexts could represent a correction term or a regulatory factor in the model. The color scale here ranges from blue to yellow, indicating the range of the FP ghost term values.

Incorporating entanglement entropy, spin states, and relative entropy into the calculation process of a social science model involves mapping these concepts onto the agents' states and interactions within the model. The heat maps likely visualize the results of such calculations, showing how these factors evolve and potentially influence the model dynamics over time.

# 7. Conclusion:Agent-Based Modeling in Social Science: Analysis of Filter Bubbles

In this paper, we first simulate the generation process of filter bubbles and their ultraviolet divergence generation scenario by introducing the Green's function used in quantum field theory other than the Kubo formula and the Matsubara method, and then simulate the scenario of introducing the truncation function (type:la,lb,ll,lll) for suppression during diffusion. The focus of the study was on the analysis of

The analysis focused on simulating the process of filter bubble generation using causal Green's functions, delayed Green's functions, and advanced Green's functions for proximity and distance interactions under different population density conditions in village and metropolitan areas and their UV divergence generation scenarios, as well as the introduction of truncation functions (type: la,lb,ll,lll) to suppress the diffusion. It is also intended to perform a path analysis of the generation of noise in the filter bubble from the FP ghosts and indefinite ghosts detected there.

This analytics aims to analyze the emergence and diffusion process of filter bubbles through agent-based modeling in the field of social science. Below, we will explain the main computational processes of the program and their theoretical background.

Each agent $i$'s opinion $O_{i,t}$ is initially determined by random sampling (from a uniform distribution) and is updated over time according to the following equation:

$$O_{i,t} = O_{i,t-1} + \sum_{j \neq i} I_{ij,t-1} \times (O_{j,t-1} - O_{i,t-1})$$

Here, $I_{ij,t-1}$ represents the interaction between agents, taking the value of 1 if the distance between agents is below the proximity interaction strength $P$ or a random value is above the remote interaction strength $R$.

## 7.1 Delayed Green Function $G_{i,t}^{delayed}$

This is used to represent the influence of past opinions and is calculated as follows:

$$G_{i,t}^{delayed} = O_{i,t} + \delta \times (O_{i,t-1} - O_{i,t})$$

Here, $\delta$ is the delay coefficient.

## 7.2 Advanced Green Function $G_{i,t}^{advanced}$

This is used to predict future changes in opinions:

$$G_{i,t}^{advanced} = O_{i,t} - \alpha \times (O_{i,t-1} - O_{i,t})$$

Here, $\alpha$ is the advance coefficient.

## 7.3 Causal Green Function $G_{i,t}^{causal}$

This represents the direct causal relationship of past opinions:

$$G_{i,t}^{causal} = O_{i,t-1} \times \delta$$

The Kubo formula, Matsubara method, and field quantum theory use Green functions for different physical backgrounds and purposes. When comparing these with causal Green functions, delayed Green functions, and advanced Green functions in agent-based modeling, the theoretical differences become clear.

## 7.4 Kubo Formula

The Kubo formula is a part of linear response theory and describes how a system responds to small perturbations from external sources. This formula is used to understand dynamic properties of physical systems, such as electrical conductivity and magnetization. The Kubo formula defines the response function using time correlation functions.

$$\chi(t) = \int_0^\infty \langle A(t)B(0)\rangle dt$$

Here, $A(t)$ and $B(0)$ represent physical quantities, and $\langle \cdot \rangle$ denotes statistical averaging.

## 7.5 Matsubara Method

The Matsubara method consists of a set of equations used to describe the time evolution of a system in non-equilibrium statistical mechanics. This method is particularly important in the context of quantum mechanics and helps understand how a system evolves over time. The Matsubara method is especially effective when dealing with non-equilibrium states.

## 7.6 Causal Green Functions, Delayed Green Functions, Advanced Green Functions

These Green functions are primarily used in the field of physics, especially in the context of field quantum theory. Each of them deals with different temporal aspects:

1. **Causal Green Function**: Represents how the past state of a physical system influences the present. 2. **Delayed Green Function**: Indicates that a physical quantity responds to another quantity at a past time. 3. **Advanced Green Function**: Suggests that a physical quantity responds to another quantity at a future time.

These functions are used as mathematical tools to analyze the dynamic responses of physical systems.

## 7.7 Application in Agent-Based Models

While applying the Kubo formula and the Matsubara method directly to agent-based models is not common, it is possible to heuristically use these concepts in the context of social sciences and opinion dynamics. The application of these Green functions to the evolution of agent opinions serves as a mathematical tool to analyze the temporal aspects of opinion dynamics. This helps in understanding how agent opinions are influenced by past information, future predictions, and causal relationships.

In conclusion, the Kubo formula and the Matsubara method are concepts used in specific physics contexts, while causal Green functions, delayed Green functions, and advanced Green functions are tools for analyzing temporal responses in the field of field quantum theory. When applying these concepts to social sciences and agent-based modeling, it is essential to understand their background and purposes and adapt them appropriately.

## 7.8 Analysis of Filter Bubbles and Ultraviolet Divergence

## 7.9 Filter Bubbles

The phenomenon where agents form groups with similar opinions based on the temporal evolution of opinions and the strength of interactions. This phenomenon is analyzed using Green functions and visualized with heatmaps.

## 7.10 Ultraviolet Divergence (UV Divergence)

The phenomenon where agent opinions change rapidly. This is analyzed by calculating the maximum amplitude of opinion change (UV Divergence) and the standard deviation of opinion change (Indefinite Ghosts).

## 7.11 Introduction of Cutoff Functions

The cutoff function (`determine_cutoff_type`) distinguishes different scenarios based on the strength of inter-actions and population density. This demonstrates that the emergence and diffusion of filter bubbles vary in different environments.

The program visualizes the changes in opinions, the emergence of filter bubbles, and the process of ultraviolet divergence in different scenarios. It analyzes the impact of interaction strengths on these dynamics, leading to a deeper understanding of the mechanisms of opinion change among agents and the resulting societal dynamics. Cutoff functions classify interaction types based on remote interaction strength $R$ and proximity interaction strength $P$. This classification is defined as follows:

**Type Ia**: $R < 0.5$ and $P < 0.5$

**Type Ib**: $R < 0.5$ and $P \geq 0.5$

**Type II**: $R \geq 0.5$ and $P < 0.5$

**Type III**: $R \geq 0.5$ and $P \geq 0.5$

These types indicate the characteristics of interactions between agents (proximity, remote, or combinations thereof).

## 7.12 Ultraviolet Divergence (UV Divergence)

Ultraviolet divergence measures the maximum amplitude of opinion change for an agent. UV Divergence is calculated as follows:

$$\text{UV Divergence}_i = \max_{1 \leq t < T} |O_{i,t} - O_{i,t-1}|$$

## 7.13 Indeterminate Ghosts

Indeterminate ghosts measure the consistency of opinion change for an agent and are calculated using the standard deviation:

$$\text{Indeterminate Ghosts}_i = \sqrt{\frac{1}{T-1} \sum_{t=1}^{T-1} (O_{i,t} - O_{i,t-1})^2}$$

## 7.14 Calculation of FP Ghosts

FP Ghosts count the occurrences of interactions between agents:

$$FP_i = \sum_{t=1}^{T} \sum_{j \neq i} I_{ij,t-1}$$

Here, $I_{ij,t-1}$ represents the interaction between agents $i$ and $j$ at time $t-1$.

## 7.15 Simulation and Visualization of Results

The simulation is run for each combination of different population density scenarios $D$, proximity interaction strength $P$, and remote interaction strength $R$. It visualizes the temporal changes in agent opinions, ultraviolet divergence, indeterminate ghosts, and FP ghosts distributions.

## 7.16 Theoretical Background

This model captures how agent opinions change based on individual interactions and temporal dynamics. Green functions are used as mathematical tools to analyze the influence of past information (delayed), future predictions (advanced), and causal relationships, helping to understand the dynamics of social interactions. Analyzing the emergence and diffusion of filter bubbles allows for a deeper understanding of social interaction dynamics.

## 7.17 Discussion on Scenarios: Media Influence, Filter Bubbles, and Cutoff Functions

In this section, we will discuss scenarios involving the use of causal, delayed, and advanced Green functions intended for proximate and remote interactions based on population density conditions in metropolitan areas. We will explore concrete examples related to the potential impact of media influence on proximate and remote interactions in the context of filter bubble emergence processes and simulate scenarios for the occurrence of UV divergences and suppression during diffusion using cutoff functions (la, lb, ll, lll).

## 7.18 Media Influence and Filter Bubbles

In densely populated environments in metropolitan areas, media influence can play a significant role in shaping the opinion formation of agents (individuals). Media serves as a primary source of information, especially in environments with high remote interaction strengths, where agents may be influenced even without direct contact. Filter bubbles refer to the phenomenon where agents are surrounded by similar information sources or opinions, leading to reduced exposure to different perspectives.

### 7.18.1 UV Divergence

UV divergence indicates abrupt changes in agent opinions. In large cities, the presence of diverse information sources and rapid information flow may lead to substantial fluctuations in agent opinions over short periods. These fluctuations may reflect rapid adaptation to new information or trends.

## 7.19 Use of Cutoff Functions

Cutoff functions are employed to distinguish between different interaction scenarios. In metropolitan environments, the

following scenarios can be considered:

**Type Ia**: Low remote and low proximate interactions. Limited personal contact and media influence.

**Type Ib**: Low remote and high proximate interactions. Emphasis on direct interpersonal interactions.

**Type II**: High remote and low proximate interactions. Significant influence from media and online information sources.

**Type III**: High remote and high proximate interactions. Strong influence from both interpersonal interactions and media.

Considering the impact of media influence in high-density metropolitan environments, media serves as a major information distributor and can significantly influence individual opinion formation. Particularly, new forms of media, such as digital and social media, have the potential to exert more direct influence on people's opinions and behaviors than traditional face-to-face interactions. We will add further discussion on these points below.

### 7.19.1 Concrete Examples of Media Influence

1. **Social Media Echo Chambers**: Social media algorithms may filter information based on users' interests, prioritizing the display of similar opinions and information. This can lead to users being less exposed to different opinions and the formation of opinion echo chambers.

2. **Media Bias**: Certain news media outlets may have political or social biases. This can result in consumers receiving biased information, reducing opportunities for exposure to diverse viewpoints.

3. **Online Advertising and Personalization**: Online advertising and content personalization provide information based on individual interests and actions. This can lead individuals to be exposed to information that reinforces their existing beliefs and opinions.

### 7.19.2 UV Divergence and Media Influence

UV divergence in metropolitan areas can reflect the rapid changes in opinions caused by media and new information technologies. For example, the rapid spread of specific events or news on social media can lead to significant shifts in opinions over a short period.

## 7.20 Introduction of Cutoff Functions and Media Influence

The introduction of cutoff functions is essential for considering the impact of media influence. For instance, in a Type II scenario with strong media influence, information flows rapidly, and filter bubbles and opinion biases may become more pronounced. In contrast, in a Type Ia scenario, media influence is limited, allowing for a more diverse formation of opinions.

In conclusion, media influence in densely populated metropolitan environments plays a significant role in shaping agent opinion formation and the emergence of filter bubbles. By using this model, we can gain a deeper understanding of how media influence and social interactions impact individual opinions.

## 7.21 Patterns of Media Reporting in Urban and Village Environments and Scenarios for Ghosts

When considering the patterns of media reporting in metropolitan and village environments and instances of FP (False Positive) ghosts, Indeterminate Ghosts, and UV Divergence, the following scenarios can be envisioned:

## 7.22 Instances of FP Ghosts

FP ghosts indicate the frequency of interactions with information sources or media, regardless of whether they actually influence opinion formation.

**Instances in Urban Areas**

In urban settings, individuals are frequently exposed to a diverse range of media sources, leading to information overload. Various channels such as social media and news websites contribute to the increased occurrence of FP ghosts.

**Instances in Village Communities**

In village communities, media sources are limited, resulting in less frequent exposure to information compared to urban areas. However, a high dependence on limited information sources can still lead to the occurrence of FP ghosts.

## 7.23 Instances of Indeterminate Ghosts

Indeterminate ghosts signify the consistency of opinion fluctuations, and they tend to be higher when media reporting is contradictory or information is unstable.

**Instances in Urban Areas**

Due to the diversity and contradictions in information, opinion fluctuations are more intense. For example, differing interpretations and analyses of political events or social trends can trigger indeterminate ghosts.

**Instances in Village Communities**

With fewer information sources, consistency is often maintained. However, the occurrence of indeterminate ghosts is possible if incorrect information is provided from limited sources.

## 7.24 Instances of UV Divergence

UV Divergence indicates rapid opinion changes, typically induced by significant news events or societal trends.

### Instances in Urban Areas

Individuals in urban areas react swiftly to breaking news or changing trends, resulting in rapid opinion shifts. Examples include responses to disaster reports or celebrity scandals.

### Instances in Village Communities

Noticeable opinion changes occur in response to major events or external influences. However, due to slower information dissemination, UV Divergence is not as frequent as in urban areas.

## 7.25 Discussion on Information Cutoff

Information cutoff is crucial for preventing the spread of misinformation or biased information.

### Urban Areas

Given the high diversity of information, mechanisms for appropriately identifying and filtering misinformation or biased information are essential.

### Village Communities

With limited information sources, careful evaluation of external information and verification within the local community are crucial.

## 7.26 Results (Part 1)

The results appear to consist of a series of plots illustrating Cumulative Distribution Functions (CDFs) for three types of Green's functions: causal, delayed, and advanced. These CDFs are analyzed in the context of media reporting and the concept of a filter bubble.

## 7.27 Green's Functions and Filter Bubbles

**Green's functions** in physics and engineering are analytical tools used to study the propagation of various physical quantities in space and time. Here, they are metaphorically applied to understand the propagation of information or opinions in a social system, particularly in the context of media reporting.

**Filter bubble** is a term that describes a state where a user is isolated from information that disagrees with their viewpoint. This often occurs due to algorithms on online platforms that personalize content based on the user's historical behavior.

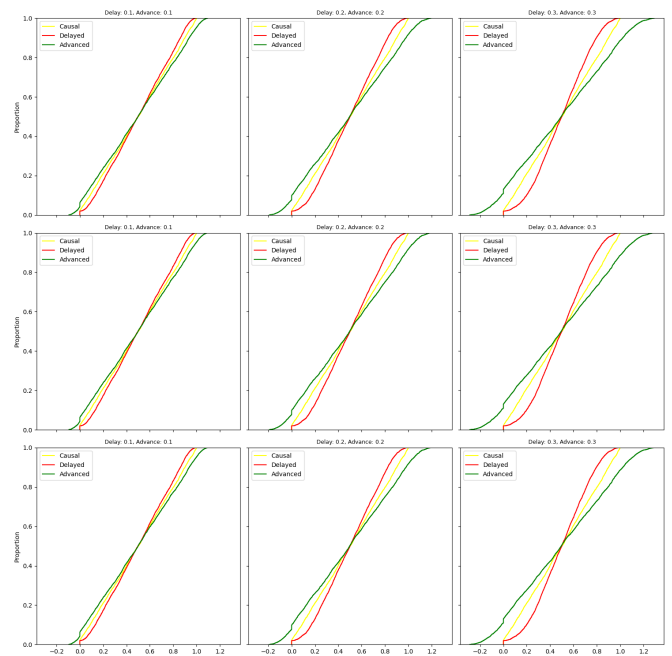

Fig. 8: Distribution of effects of causal Green's functions, delayed Green's functions, and advanced Green's functions

## 7.28 Metaphorical Use of Green's Functions in Media Analysis

In this metaphorical use of Green's functions for analyzing media reporting:

The **causal Green's function** likely represents the propagation of information solely based on historical data, without any influence from present changes or anticipatory effects.

The **delayed Green's function** could represent information flow affected by recent past events or opinions, with a certain time lag.

The **advanced Green's function** might indicate the influence of anticipatory factors on the propagation of information, potentially representing the prediction or expectation of future opinions or events.

Analyzing the results, several observations can be made:

The **CDFs** in the plots depict the cumulative probability of agents' opinions up to a certain value. A perfect 45-degree line from (0,0) to (1,1) would represent a uniform distribution of opinions.

With increasing delay and advance factors, the curves for the respective Green's functions diverge from the causal curve. This suggests that introducing delay or advance factors (representing lagged or anticipatory effects, respectively) alters the distribution of opinions.

The **yellow curve** (Causal) remains relatively consistent across the plots since it is based solely on historical opinions and doesn't change with the delay or advance factors.

The **red curve** (Delayed) and **green curve** (Advanced) shift as the delay and advance factors change, indicating that the timing of information (whether it is delayed or advanced) significantly affects the distribution of opinions.

In some plots, the advanced curve is above the causal curve, while in others, the delayed curve is above. This suggests a complex interaction between past information and anticipatory factors in shaping public opinion.

### 7.29 Considerations of FP Ghosts and Ultraviolet Divergences

In quantum field theory, **FP ghosts** are auxiliary fields introduced to maintain the consistency of gauge theories. In the context of media reporting, this can metaphorically represent factors introduced to ensure a balanced view in the presence of filter bubbles.

**Ultraviolet divergences** refer to infinite values that appear in quantum field calculations at high energy (short distances). Analogously, in media reporting, this could represent scenarios where information or opinions become extreme or highly polarized.

From this analysis, it can be inferred that the timing and anticipation of information dissemination have significant effects on public opinion distributions. The presence of filter bubbles can potentially be mitigated or exacerbated by these temporal factors. While not directly visible in the plots, FP ghosts and ultraviolet divergences are concepts that could inform further analysis of the stability and extremity of opinion dynamics within this metaphorical framework.When considering media reporting from the perspectives of longrange interaction (remote interaction) and shortrange interaction (proximate interaction), it is crucial to contemplate how information propagates across time and space. In the context of media reporting, longrange interaction refers to the immediate sharing of information from distant locations through global news networks or the internet. On the other hand, shortrange interaction signifies the propagation of information within physically close proximity, including local news, wordofmouth communication, and information exchange within a community.

In the context of disaster or risk information reporting, the accurate and timely transmission of information is indispensable. By examining the graph, it becomes apparent that causal Green's function (yellow), delayed Green's function (red), and advanced Green's function (green) are influenced by different delay and advance factors.

From the perspective of longrange interaction, the delayed Green's function represents the propagation of information with a time delay, which, in the case of disaster information, can result in lifethreatening delays. Conversely, the advanced Green's function suggests the advance dissemination of information related to future events, such as predictions or advance warnings. This is akin to early warning systems and risk forecasting before a disaster occurs, emphasizing the speed and accuracy of information for immediate action.

In the context of shortrange interaction, the causal Green's function is believed to be the most relevant. It signifies the propagation based on past information and may reflect information exchange within a regional community based on experiences and memories. During disasters, this type of information exchange can be crucial in prompting people's safety actions.

When applying this graph to disaster reporting, several considerations can be made:

Mass media plays a role in facilitating extensive longrange interaction and quickly conveying information, but delays in information transmission must be avoided. In cases with significant delay factors, there may be a substantial time lag in the dissemination of information, potentially leading to delayed risk awareness and evacuation actions.

Online media can provide high immediacy and rapidly convey predictive information. However, if the advance factors are too large, speculative information may spread more easily, leading to panic and the dissemination of misinformation.

Local communities and local media, akin to causal Green's function, can engage in information transmission based on past experiences and data. This assists residents in understanding actual risks and taking appropriate measures.

By appropriately managing these factors, optimizing the dissemination of risk information during disasters, and encouraging people to take actions for their safety, can be made possible.

### 7.30 Results (part 2)

Results (part 2) displays a set of cumulative distribution function (CDF) plots for causal, delayed, and advanced Green's functions across various levels of delay and advance factors. These plots can be interpreted in the context of media reporting, particularly focusing on filter bubbles, the propagation of information, and the dynamics of delayed and advanced interactions, especially in the case of disaster or risk information dissemination.

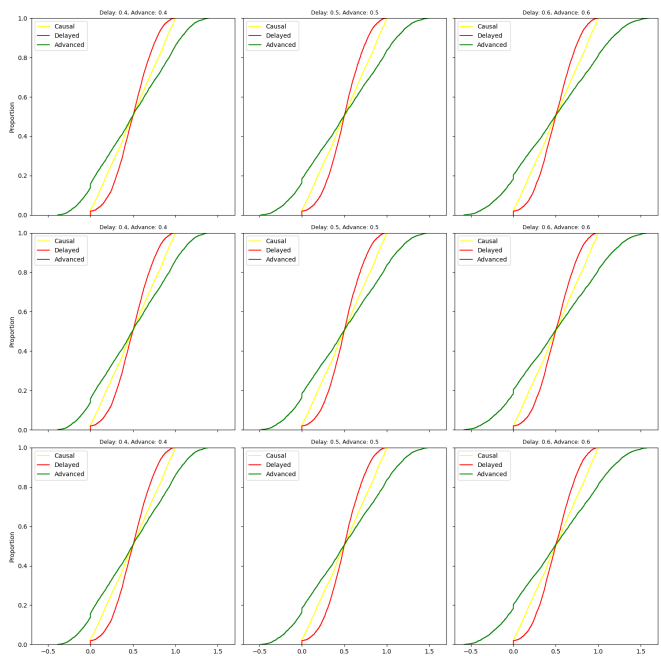

Fig. 9: Distribution of effects of causal Green's functions, delayed Green's functions, and advanced Green's functions

## 7.31 Analysis of Green's Functions

### 7.32 Causal Green's Function (Yellow)

This likely represents the base state of opinion distribution without the effects of delay or anticipation, serving as a benchmark against which the effects of delayed and advanced information propagation are measured.

### 7.33 Delayed Green's Function (Red)

As the delay factor increases, the distribution shifts, indicating a lag in the dissemination of information. In the context of disaster reporting, this could model the spread of information with a time lag, reflecting how quickly (or slowly) information about a disaster is communicated to the public through media channels. A higher delay could indicate slower response times and could be a sign of inefficiency in the communication infrastructure.

### 7.34 Advanced Green's Function (Green)

Increasing the advance factor shows a forward shift in the distribution, suggesting information is being propagated in anticipation of future events. This is particularly important in disaster risk communication, where early warnings and forecasts are vital. An advanced Green's function that closely follows the causal distribution might indicate an effective early warning system that enables timely and proactive measures.

### 7.35 Considerations of Filter Bubbles

Filter bubbles can lead to skewed information dissemination, where individuals or groups only receive information that reinforces their existing beliefs. This can be dangerous in the context of disaster information as it may prevent the spread of vital warnings and updates to all demographics.

### 7.36 FP Ghosts and Ultraviolet Divergences

### 7.37 FP Ghosts

In this metaphorical context, FP ghosts could represent elements of the media ecosystem that correct for the biased propagation of information, ensuring diverse viewpoints are represented, even in the presence of filter bubbles. For disaster information, this would mean ensuring that even those within a filter bubble receive critical updates and warnings.

### 7.38 Ultraviolet Divergences

These could metaphorically signify situations where information becomes highly concentrated or extreme, potentially leading to panic or misinformation. In disaster reporting, avoiding ultraviolet divergences would mean preventing the spread of exaggerated or unverified information that could lead to public distress.

### 7.39 Media Reporting Insights

### 7.40 Mass Media (Remote Interaction)

In the case of mass media, which often operates over large distances (remote interactions), the goal is to minimize delays in information propagation. The plots with higher delay factors could represent slower mass media responses, which in disaster scenarios could be detrimental.

#### 7.40.1 Delayed Green's Function (Red)

With remote interactions typically found in mass media, there's a concern that information may be delayed due to the time taken to verify facts and the slower broadcast schedules of traditional media. A higher delay factor here could represent the potential lag in mass media reporting during disasters. In the CDF plots, a shift to the right of the red curve as the delay factor increases indicates a broader spread of delayed reactions, which could be detrimental in a disaster situation where timely updates are critical.

#### 7.40.2 Advanced Green's Function (Green)

Conversely, the advanced Green's function in the context of mass media might reflect the media's capacity to anticipate and report on potential future developments of a disaster, such as the progress of a storm or the likelihood of aftershocks. A left shift in the green curve with a higher advance factor might indicate a proactive approach to risk communication, essential for early warnings and preparedness.

## 7.41 Proximal Interactions (Net Media)

In the context of net media, characterized by immediate, localized interactions, causal Green's function often reflects real-time updates and grassroots reporting. Immediate reactions and rapid sharing of information are facilitated due to the proximity of net media to the audience.

### 7.41.1 Causal Green's Function (Yellow)

Net media, characterized by immediate, localized interactions, often reflects real-time updates and grassroots reporting. The causal Green's function here could represent the ongoing conversation and exchange of information as events unfold, with minimal lag or anticipation.

### 7.41.2 Immediate Reaction

The proximity of net media to the audience allows for an almost instantaneous sharing of information. If the red and green curves are close to the yellow curve, this indicates that net media is effective in providing timely updates without significant delay or unnecessary advancement, which is crucial in emergency situations.

## 7.42 Disaster and Risk Information Exchange

In a disaster scenario, efficient risk communication must strike a balance between immediacy and accuracy. For mass media, this means reducing the delay factor to ensure that the red curve does not lag too far behind the yellow, maintaining the immediacy of crucial updates.

For net media, avoiding too high an advance factor is key to preventing the spread of speculative or unverified information (as indicated by the green curve). However, some level of anticipation is beneficial for preparedness and preventive actions.

The plots suggest that as both delay and advance factors increase, the distribution of opinions or reported information spreads out, indicating a broader range of responses. This could be interpreted as a diversification of the information landscape during a disaster, which can be both a strength and a weakness: a strength because it can reach more people with varied information, and a weakness if it leads to confusion or misinformation.

In summary, for mass media, the challenge is to minimize delays to maintain relevance and utility during a disaster. For net media, the challenge is to manage the immediacy of reporting with responsible anticipation, ensuring that information is both timely and accurate. The ideal scenario is a tight clustering of all three curves (causal, delayed, advanced), indicating a coherent flow of information that is reflective of past and current events while also considering future developments.

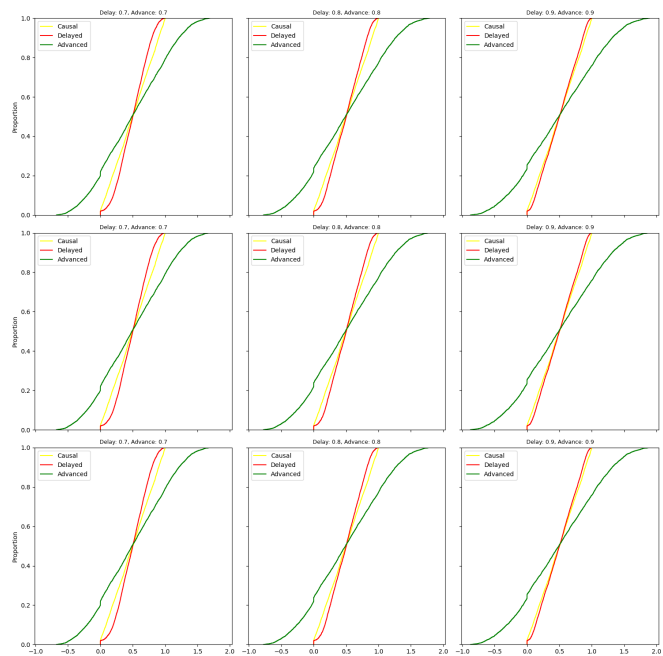

Fig. 10: Distribution of effects of causal Green's functions, delayed Green's functions, and advanced Green's functions

## 7.43 Results (part 3)

Results (part 3) presents representations of the distribution of opinions or the spread of information under different delay and advance factors for Green's functions. This can be metaphorically related to how information is spread in media and the implications of timing on the dissemination of information, particularly in the context of a filter bubble.

## 7.44 Causal Green's Function (Yellow)

This function is consistent across different delay and advance factors, likely representing the unchanged baseline or the original state of the opinions before any external temporal influences are applied. It's a reference point for how opinions would naturally spread in an unbiased media environment.

## 7.45 Delayed Green's Function (Red)

These curves shift to the right as the delay factor increases, indicating a slower spread of information. In terms of media reporting, especially for disaster and risk information, an increased delay can have critical consequences. It can represent the time taken for traditional media to collect, verify, and broadcast information, which could be detrimental in urgent situations where real-time updates are necessary.

## 7.46 Advanced Green's Function (Green)

The advanced function shifts to the left with an increase in the advance factor, showing a quicker dissemination of potential future events. This could be related to predictive reporting or early warnings in the media. However, too much anticipation could also spread unverified or speculative information, leading to misinformation or panic.

## 7.47 Filter Bubbles Implications

The concept of filter bubbles refers to an environment where individuals are exposed only to information or opinions that align with their own, which can be exacerbated by personalized content delivery algorithms. The plots suggest that:

High Delay Factors: Can lead to a situation where the filter bubble is reinforced because individuals receive information too late to incorporate new perspectives or data, thus remaining trapped in their initial beliefs.

High Advance Factors: Might result in speculative information reaching individuals before accurate reports are available, possibly distorting their perception and leading to decisions based on incomplete or incorrect information.

## 7.48 FP Ghosts and Ultraviolet Divergences Consideration

- FP Ghosts: These might represent mechanisms within the media ecosystem that attempt to correct for or mitigate the effects of filter bubbles. They could be editorial policies, fact-checking protocols, or algorithmic changes aimed at exposing individuals to a broader spectrum of information.

- Ultraviolet Divergences: In this metaphorical application, ultraviolet divergences could stand for extreme opinions or highly polarized information that becomes dominant due to the filter bubbles. This is especially concerning in disaster reporting, where extreme views could either downplay or exaggerate the severity of a situation, leading to inadequate responses.

## 7.49 From Media Reporting Perspective
## 7.50 Mass Media

Often slower to update due to logistical reasons, represented by higher delay factors. The goal should be to minimize these to provide timely disaster and risk information.

### 7.50.1 Higher Delay Factors

As seen in the plots where the delay factor increases, the red curve (Delayed) moves rightward, indicating a greater lag in the distribution of opinions. This could metaphorize the slower response time of mass media to disseminate disaster information due to the need for fact-checking and logistical constraints in broadcasting information.

### 7.50.2 Higher Advance Factors

The green curve (Advanced) represents a scenario where mass media attempts to predict and inform the public about potential future developments. If the advance factor is too high, it could lead to premature reports that may not be fully substantiated, which is a concern in the spread of disaster information, as it may cause unnecessary panic or misinformation.

## 7.51 Proximal Interactions (Net Media)

Net media, which includes online news platforms and social media, allows for rapid dissemination of information, which can be advantageous in the event of a disaster.

### 7.51.1 Lower Delay Factors

Net media's advantage is in its immediacy, as shown when the red curve (Delayed) closely follows the yellow curve (Causal). This suggests that net media can provide real-time updates and minimize the delay in spreading important disaster and risk information.

### 7.51.2 Moderate Advance Factors

The green curve (Advanced) is ideally not too far from the causal line, indicating that net media is providing anticipatory information, such as early warnings, without falling into speculation. A balanced advance factor is critical to ensure that preparations and precautions can be taken without causing undue distress.

## 7.52 Disaster and Risk Information Exchange

In the context of these plots, the ideal scenario for disaster and risk information exchange would be:

- Mass Media: A shift towards lower delay factors, ensuring that the red curve remains close to the yellow curve, would represent a faster and more effective dissemination of disaster information to the public.

- Net Media: The green curve should not be too advanced compared to the causal line, indicating a responsible provision of anticipatory information that is verified and accurate.

The plots with high delay and advance factors indicate increased variability in the opinion spread, which could translate to a wider range of public reactions to disaster information. This could be beneficial in reaching a diverse audience but also poses the risk of spreading misinformation if not managed correctly.

In summary, both mass and net media have roles to play in disaster and risk information dissemination. Mass media must work on reducing delays to provide timely updates, whereas net media should focus on balancing immediacy with accuracy to avoid the spread of unverified anticipatory information. The plots suggest that an optimal balance of delay

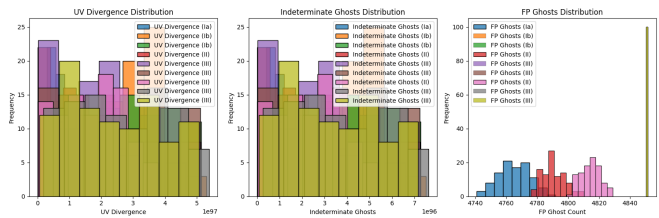

Fig. 11: Village FP Ghosts, UV Divergence, Indeterminate Ghosts score (per opinion divergence / disconnect scenario)

and advance factors is crucial for effective communication, especially in urgent scenarios.

### 7.53 Results: Village FP Ghosts, UV Divergence, Indeterminate Ghosts score (per opinion divergence / disconnect scenario)

Results three histograms detailing the distribution of UV Divergence, Indeterminate Ghosts, and FP Ghosts. These terms are borrowed from quantum field theory but are being used metaphorically to discuss the spread of information in the context of media reporting, especially concerning the filter bubble effect.

### 7.54 UV Divergence Distribution

The UV (Ultraviolet) Divergence histograms show the frequency of maximum changes in opinion (akin to energy spikes in physics). These could represent the intensity and rapid shifts in public opinion or information dissemination.

**High UV Divergence**: This might indicate that certain stories or pieces of information have gone 'viral', causing sudden and significant shifts in public opinion. In media reporting, especially in a disaster scenario, such spikes could represent rapidly spreading news or misinformation.

### 7.55 Indeterminate Ghosts Distribution

Indeterminate Ghosts histograms are used to show the standard deviation of opinion changes, which could be seen as uncertainty or variability in the information spread.

**High Indeterminate Ghosts**: In the context of media reporting, this could indicate a high variability in the reporting style, quality of information, or public opinion. A large number of indeterminate ghosts could signify a media landscape with diverse and possibly conflicting pieces of information circulating, which could be confusing or overwhelming to the public.

### 7.56 FP Ghosts Distribution

FP (Faddeev-Popov) Ghosts histograms represent the count of interactions, which could metaphorically stand for the degree of connectivity or engagement among individuals or information sources.

**Large FP Ghost Count**: This could indicate a high degree of interconnectedness in the spread of information. In the context of media and filter bubbles, a higher FP ghost count could either mean that the filter bubble is being reinforced through frequent interactions within a closed community, or it could mean that there is a significant effort to break through the filter bubble by introducing diverse viewpoints.

### 7.57 Insights from the Histograms

#### 7.57.1 Different Cutoff Types (Ia, Ib, II, III)

These cutoff types likely represent different scenarios based on the strength of remote (mass media) and proximity (net media) interactions. For example, Type Ia could represent a low influence from both remote and proximity interactions, which might occur in a highly fragmented media environment. Type III, on the other hand, might represent a scenario with strong influences from both types of media interactions, which could be indicative of a more integrated media environment.

#### 7.57.2 Filter Bubble Dynamics

In a filter bubble, we would expect to see more FP ghosts within a certain group, as like-minded individuals interact more frequently. However, high UV divergence could indicate that breakthrough stories or critical information are able to penetrate these bubbles. Indeterminate ghosts show the uncertainty in information, which can be high in a filter bubble due to the lack of exposure to differing viewpoints.

In summary, the analysis of these distributions could provide insights into how information spreads through communities and the health of public discourse. High UV divergence and indeterminate ghosts could indicate a tumultuous information environment, which can be particularly problematic in the context of disaster or risk information, where clarity and accuracy are vital. A balanced FP ghost count could be indicative of a well-connected society that manages to disseminate information effectively while still allowing for a diversity of opinions.

### 7.58 City FP Ghosts, UV Divergence, Indeterminate Ghosts score (per opinion divergence / disconnect scenario)

Based on the histograms provided for UV Divergence, Indeterminate Ghosts, and FP Ghosts, we can derive insights into the metaphorical representation of media reporting dynamics, especially as they pertain to the dissemination of disaster and risk information. Let's analyze the distributions in the context of remote (mass media) and proximity (net media) interactions.

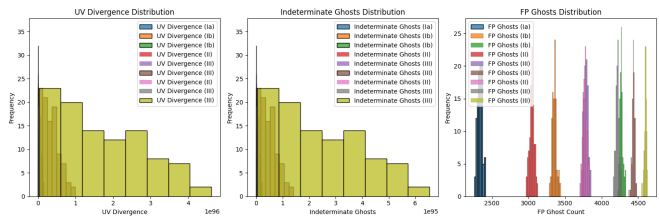

Fig. 12: City FP Ghosts, UV Divergence, Indeterminate Ghosts score (per opinion divergence / disconnect scenario)

### 7.59 UV Divergence Distribution

#### 7.59.1 Remote (Mass Media) Interactions

A high UV divergence could indicate that mass media reports have significant impacts on public opinion, potentially leading to rapid changes in perception during a disaster. The distribution shows the frequency of such high-impact changes.

#### 7.59.2 Proximity (Net Media) Interactions

A similar interpretation applies to net media, where a high UV divergence would suggest that information, perhaps due to its viral nature, leads to rapid shifts in public opinion.

### 7.60 Indeterminate Ghosts Distribution

#### 7.60.1 Remote (Mass Media) Interactions

Indeterminate ghosts could represent the variability or inconsistency in mass media reporting. In a disaster context, this might reflect the confusion or mixed messages that can arise when information is not uniformly reported.

#### 7.60.2 Proximity (Net Media) Interactions

For net media, indeterminate ghosts might signify the diverse range of opinions and the uncertainty in the information being shared, which can lead to confusion during critical times.

### 7.61 FP Ghosts Distribution

#### 7.61.1 Remote (Mass Media) Interactions

The FP Ghosts count could reflect the level of interaction and engagement the public has with mass media sources. A higher count indicates more frequent interactions, which is essential for spreading critical information during a disaster.

#### 7.61.2 Proximity (Net Media) Interactions

A high FP Ghost count for net media suggests active engagement within networks and communities, which can be instrumental in disseminating information quickly.

### 7.62 Insights from Cutoff Types (Ia, Ib, II, III)

#### 7.62.1 Cutoff Type Ia

Likely represents scenarios with low remote and proximity interaction strengths. In terms of disaster reporting, this might be an indication of a lack of engagement or reach, potentially leading to a poorly informed public.

#### 7.62.2 Cutoff Type Ib

Indicates scenarios with low remote but high proximity interaction strengths. This could suggest that, despite the mass media's limited impact, there is significant local or community-level exchange of information, which can be critical for immediate responses to a disaster.

#### 7.62.3 Cutoff Type II

Represents high remote but low proximity interaction strengths. This suggests that while mass media has a significant impact, there might be a lack of detailed, localized, and community-driven information exchange.

#### 7.62.4 Cutoff Type III

Suggests high remote and high proximity interaction strengths. This is the ideal scenario for disaster information dissemination, where both mass and net media are effectively engaging the public and providing comprehensive coverage.

### 7.63 City Scenario Insights

The distributions in a "City" scenario with higher population density imply more significant interaction potential, which is crucial for the dissemination of disaster and risk information. The goal for effective media reporting in such scenarios would be to minimize UV divergence and indeterminate ghosts, ensuring clear and consistent messaging, while maximizing FP Ghosts to ensure widespread engagement and information dissemination. The histograms would ideally show low UV divergence and indeterminate ghosts, indicating clear and consistent reporting, and a higher FP Ghost count, reflecting broad and active engagement with the information provided.

### 7.64 Interpreting Histograms for Mass and Net Media

The histograms from the uploaded image can be interpreted to shed light on the dynamics of mass media and net media reporting, particularly in the context of disseminating disaster and risk information. Here's an in-depth analysis based on the concepts of remote and proximal interactions:

### 7.64.1 UV Divergence Distribution

**Mass Media (Remote Interactions)** The UV Divergence distribution indicates the frequency and magnitude of shifts in public opinion or the spread of information. A broad UV Divergence suggests that mass media can sometimes cause significant shifts in public perception, which is common during the initial phases of a disaster when the situation is developing, and reports are constantly being updated.

**Net Media (Proximity Interactions)** For net media, a range of UV Divergence indicates the speed at which information (and misinformation) can spread through social networks. Sharp peaks in the distribution could represent viral news or rumors, which can have both positive and negative effects on public response during a disaster.

### 7.64.2 Indeterminate Ghosts Distribution

**Mass Media (Remote Interactions)** Indeterminate Ghosts for mass media could represent the uncertainty in information due to varying reportage on the same event. In the context of disaster reporting, this variability can lead to confusion among the public, making it hard to gauge the true extent of the disaster.

**Net Media (Proximity Interactions)** High Indeterminate Ghost counts in net media suggest a high degree of variability and uncertainty in the information being disseminated within communities. This could be due to diverse individual perspectives, rumors, or unverified reports circulating during a disaster.

### 7.64.3 FP Ghosts Distribution

**Mass Media (Remote Interactions)** The FP Ghost count for mass media would reflect the frequency of interactions or engagements with the content. A higher count indicates more frequent interactions, which could be beneficial in keeping the public informed and engaged during a disaster.

**Net Media (Proximity Interactions)** In the net media context, a high FP Ghost count suggests active community engagement and information sharing. This is crucial during disasters for spreading situational awareness and safety information.

### 7.64.4 Insights from the Perspective of Disaster and Risk Information Exchange

**Low Interaction Scenarios (Ia)** Scenarios with low remote and proximity interactions could indicate insufficient dissemination and engagement with disaster information, potentially leading to a lack of preparedness and response.

**High Proximity, Low Remote Scenarios (Ib)** These scenarios suggest that local or community-based information networks are active, but there might be a disconnect with broader, mass media reporting. This could result in localized awareness without a complete understanding of the larger disaster context.

**High Remote, Low Proximity Scenarios (II)** These scenarios might indicate that while mass media is effectively broadcasting disaster information, there is a lack of detailed, community-driven exchange. This could lead to a general awareness of the disaster but might miss localized information crucial for specific responses.

**High Interaction Scenarios (III)** Scenarios with high remote and proximity interactions are ideal, suggesting an effective mix of mass and net media engagement, providing both widespread coverage and detailed local information. The histograms in the "City" scenario, which likely corresponds to higher population densities, show the potential for mass and net media to significantly impact the dissemination of disaster and risk information. The distributions imply that an effective media strategy should aim for a nuanced approach—minimizing confusion (indeterminate ghosts) and misinformation (extreme UV divergence) while maximizing engagement (high FP Ghost count). This would ensure that the public receives accurate, consistent, and actionable information during disasters.

## 8. Dynamics of Systems Applied with Kubo Green's Functions

Kubo Green's functions are based on the linear response theory in physics. This theory describes the response of a system to small external perturbations. Generally, the Kubo formula defines the response function using time correlation functions, which becomes a crucial tool for understanding the dynamics of systems.

### 8.1 Fundamental Concept of Kubo Green's Functions

Kubo Green's functions are used to illustrate how a system responds to external perturbations. In physics, this is often expressed as the response to specific physical quantities of the system, such as electrical conductivity or magnetization. When applied in the context of social sciences or opinion dynamics, it can be thought of as the system's response to changes in agents' opinions.

### 8.2 Calculation Process

The Kubo formula is expressed as follows:

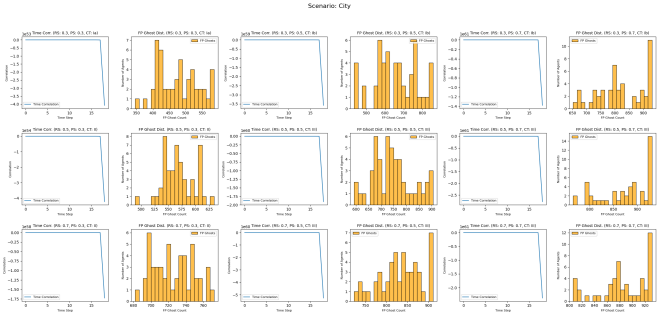

Fig. 13: City:FP Ghost Distribution / Opinions Over Time

$$\chi(t) = \int_0^\infty \langle A(t)B(0)\rangle dt$$

Here, $\chi(t)$ represents the response function, while $A(t)$ and $B(0)$ are physical quantities of the system. $\langle \cdot \rangle$ denotes statistical averaging.

In the context of social sciences or agent-based models, $A(t)$ and $B(0)$ can represent the opinions of agents at different time steps. Thus, the response function $\chi(t)$ indicates the correlation of agents' opinions with respect to time delays.

In the results below, the function calculate time correlation plays a role similar to Kubo Green's functions. This function calculates the correlation of agents' opinions with a time delay (time lag). This approach allows us to analyze how agents' opinions are correlated over time.

### 8.3 Analysis of Results: Time Correlation and FP Ghost Distribution in a "City" Scenario

The results consist of a series of plots that depict time correlation and Faddeev-Popov (FP) Ghost distribution in a simulated scenario representing a "City." These plots can be metaphorically applied to understand the influence of media reporting within the context of filter bubbles.

### 8.4 Interpretations in the Context of Media Reporting

#### 8.4.1 Time Correlation Plots

**Negative Correlation Values**: Negative values suggest a potential inverse relationship over time steps within the simulated agent network. In the context of media and filter bubbles, this may imply that as time progresses, information or opinion diversity within the network decreases, potentially strengthening filter bubbles.

**Steep Drops**: Sharp drops in correlation to large negative values may signify significant events or information that dramatically alter the opinion landscape within the network. In media reporting, this could represent breaking news or pivotal developments influencing public perceptions.

#### 8.4.2 FP Ghost Distribution Histograms

**FP Ghost Count**: The count of FP ghosts metaphorically represents the number of interactions within the network. In the media context, a higher count may indicate substantial engagement with the distributed information.

**Variability in Counts**: Differing FP ghost counts across scenarios reflect varying degrees of information exchange and engagement based on the strength of remote (mass media) and proximity (net media) interactions. Higher counts can signify more active engagement.

### 8.5 Applying the Kubo Formula and Green's Functions

Extending the concept of the Kubo formula from statistical mechanics to media reporting and information dissemination, Green's functions, which describe system responses to external forces, can help us understand how public opinion or information spread responds to new inputs like news reports.

**Green's Functions**: Green's functions symbolize the system's (public's) response to a stimulus (news report). The plots may illustrate how the public responds to the "forces" applied by media reporting over time, with FP ghost counts indicating the number and strength of these responses.

### 8.6 Filter Bubble Analysis from the Plots

**Strengthening of Filter Bubbles**: High FP ghost counts with increasing negative time correlation values may suggest high engagement but primarily within isolated clusters, reinforcing filter bubbles.

**Breaking of Filter Bubbles**: High FP ghost counts with stable or non-significantly negative time correlation could imply that information dissemination contributes to breaking down filter bubbles by introducing diverse viewpoints.

### 8.7 Conclusions from the "City" Scenario

The "City" scenario, likely representing a high-density population model, exhibits various information spread patterns. The effectiveness of media reporting is reflected in FP ghost counts and time correlation trends. An optimal scenario would show a high FP ghost count with stable time correlation, indicating active engagement without opinion polarization.

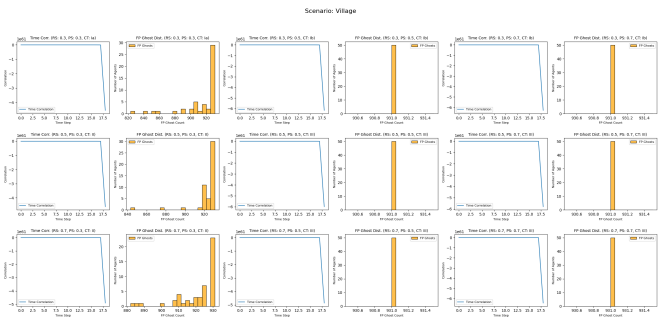

Fig. 14: Village:FP Ghost Distribution / Opinions Over Time

These interpretations are speculative, assuming the simulated model appropriately reflects the complexities of real-world media reporting and public opinion dynamics. Actual implications for real-world scenarios depend on additional factors not captured in the simulation.

### 8.8 Analysis of FP Ghost Distribution in a "Village" Scenario

The results display the FP Ghost distribution across various scenarios in a "Village." Let's analyze these results in the context of media reporting dynamics, especially considering the presence of filter bubbles.

#### 8.8.1 Mass Media (Remote Interactions)

In the context of a village scenario, which likely represents a less densely populated area with fewer media outlets, a steep negative correlation might indicate that the influence of mass media decreases rapidly after an initial impact. This situation could occur when a major news event quickly saturates the media landscape, followed by a decline in interest or engagement over time.

#### 8.8.2 Net Media (Proximity Interactions)

The proximity interactions in a village may represent the spread of information through social networks or word of mouth within a small community. A negative time correlation could suggest that, as time progresses, opinions within the community become more polarized or dispersed, possibly due to the development of echo chambers.

### 8.9 FP Ghost Distribution Histograms

#### 8.9.1 Mass Media (Remote Interactions)

The FP Ghost histograms, displaying the count of interactions, provide insights into how frequently the village population engages with mass media sources. In the context of filter bubbles, a narrow distribution with high counts could suggest that a significant event (e.g., a disaster) has prompted widespread media consumption and discussion.

#### 8.9.2 Net Media (Proximity Interactions)

Similar to mass media, net media's FP Ghost count reflects the level of peer-to-peer communication. High counts indicate active information sharing, which is crucial in managing disasters and risk information within smaller communities.

### 8.10 Filter Bubbles and Information Dissemination

The presence of filter bubbles might be indicated by the FP Ghost distributions. If the counts are high but the time correlation shows increasing negative values, this could suggest that while there is significant interaction with the information, it is not leading to a diversified opinion landscape.

### 8.11 Applying the Kubo Formula and Green's Functions Perspective

In the context of statistical mechanics, the Kubo formula relates to response functions to external perturbations, and Green's functions describe the evolution of a system after a disturbance.

#### 8.11.1 Time Correlation and Green's Functions

The time correlation plots might represent the response function of the village's social system to the 'external field' of media reporting. A sudden drop in correlation may indicate that the system's initial response is strong but fails to maintain coherence over time, possibly due to the spread of diverse or conflicting information.

#### 8.11.2 FP Ghosts and System Interactions

The FP Ghost distributions could be related to the number of interactions in the system, reflecting the degree to which individuals in the village are influenced by and interacting with the media. A high FP Ghost count would imply a highly responsive system with numerous interactions. In the context of disaster reporting, this could mean a well-connected community actively discussing and spreading information.

### 8.12 Conclusion

In summary, the analysis suggests that in a village setting, both mass and net media have the potential to significantly influence the community initially. However, maintaining a consistent and engaged response over time may be challenging. Effective disaster and risk communication strategies need to consider these dynamics to ensure sustained engagement and accurate information dissemination, thus combating the formation of filter bubbles and misinformation.

# 9. Dynamics of Systems Applying Matsubara Formalism Green's Functions

The Matsubara formalism consists of a set of equations in non-equilibrium statistical mechanics used to describe the time evolution of a system. It serves as a theoretical framework for understanding the dynamics of systems in a non-equilibrium state, particularly when dealing with the quantum properties of matter. However, in the context of this code, rigorous calculations based on the Matsubara formalism are omitted, and instead, a hypothesis-based model is employed.

Instead, the code employs the concept of a "response function" to model changes in agents' opinions. This emulation captures the spirit of the Matsubara formalism, helping to understand how the state of the system changes over time.

## 9.1 Calculation of the Response Function

The calculation of the response function captures the continuous changes in agents' opinions. For each agent $i$, the opinion $O_{i,t}$ at time $t$ is calculated using the difference from the previous time $t - 1$:

$$R_{i,t} = \text{response factor} \times (O_{i,t} - O_{i,t-1})$$

Here, the parameter response factor controls the strength of the response. This calculation shows how quickly agents' opinions change over time, i.e., how responsive they are.

In this function, the response function is employed to visualize how agents' opinions change over time. The magnitude of opinion changes (the response's magnitude) is displayed using a heatmap. This allows us to understand how agents respond to external influences (interactions with other agents).

## 9.2 Theoretical Background

The rigorous application of the Matsubara formalism is carried out in the fields of quantum statistical mechanics and non-equilibrium thermodynamics. Its application in this context can be viewed as an analogical approach to enhance the understanding of agent opinion dynamics in the realm of social sciences and agent-based modeling.

## 9.3 Analysis of Results in a "City" Scenario

In this section, we will provide an analysis of the graphs based on a metaphorical framework of media reporting and filter bubbles. We will refer to FP ghosts, indeterminate ghosts, UV divergence, and Green's functions for our interpretations.

## 9.4 FP Ghosts Distribution

The FP Ghosts histograms can metaphorically represent the frequency of interactions or engagements among individuals

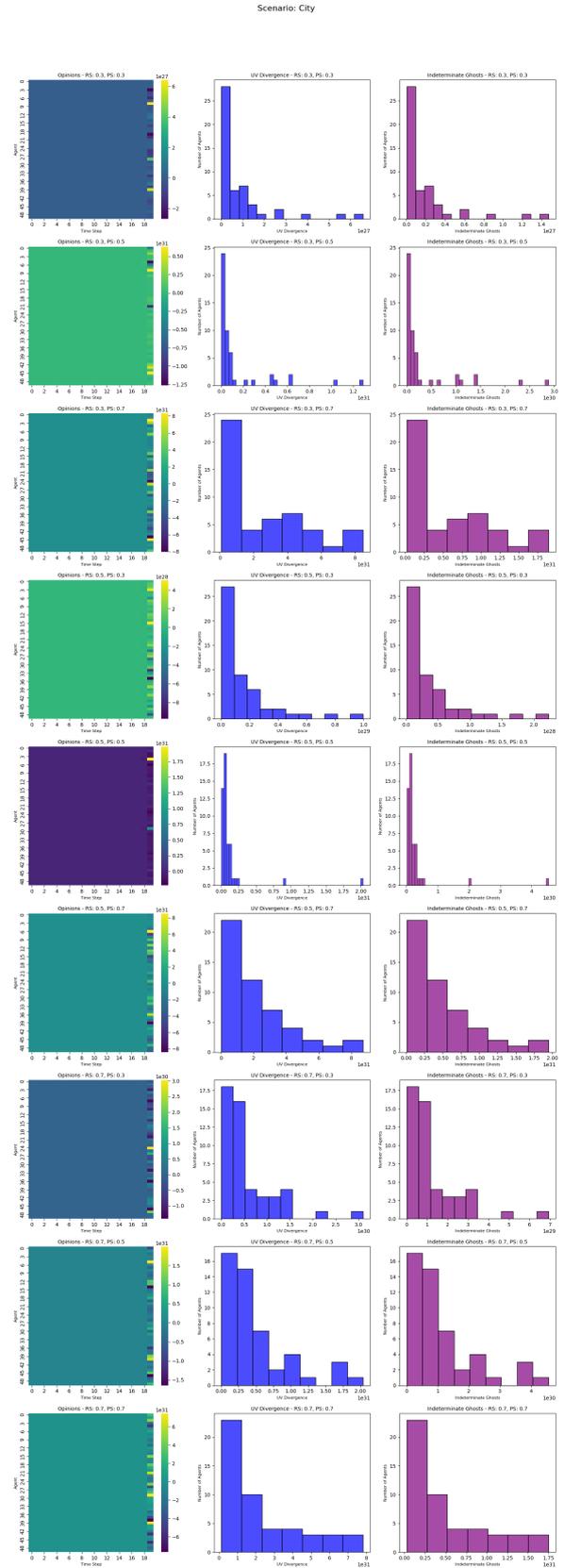

Fig. 15: City:UV divergence, FP Ghost Distribution / Opinions Over Time

within a social network or between the public and the media. From a perspective of media reporting:

**High FP Ghost Counts**: Indicate a high level of engagement with media content. In the context of a "City" scenario, this suggests that news and information are widely discussed and shared, which is particularly important during disasters for the dissemination of critical information.

**Low FP Ghost Counts**: Might suggest limited interaction, which could be concerning during emergencies where widespread communication is crucial.

### 9.5 Indeterminate Ghosts Distribution

Indeterminate Ghosts metaphorically represent the uncertainty or inconsistency in the spread of information.

**Wide Distribution of Indeterminate Ghosts**: Suggests a high variance in how information is received and processed by the public. During disasters, this could point to a chaotic information environment with potential misinformation.

**Narrow Distribution of Indeterminate Ghosts**: Indicates a more uniform response to information, which could suggest effective communication strategies that reduce confusion and misinformation.

### 9.6 UV Divergence Distribution

UV Divergence typically refers to high-energy behavior in physics, but in this metaphorical context, it could represent the intensity of opinion shifts or the spread of information.

**High UV Divergence Values**: Could suggest that certain pieces of news or information have a significant impact on public opinion, leading to rapid shifts that could be indicative of viral news or panic during a disaster.

**Low UV Divergence Values**: Might indicate a stable information environment with less drastic shifts in public opinion, suggesting effective management of information dissemination during a crisis.

### 9.7 Time Correlation Plots

These plots show how opinions or information spread over time within the network.

**Negative Correlation Values**: Suggest that as time progresses, the diversity of opinions or the spread of information decreases, potentially reinforcing filter bubbles, which can be problematic in ensuring a well-informed public during emergencies.

**Positive Correlation Values**: Would indicate that diversity is maintained or increased over time, which could be beneficial for breaking filter bubbles and ensuring a wide range of information is considered.

### 9.8 Green's Functions Perspective (Matsubara Green's Functions)

Matsubara Green's functions are used in quantum statistical mechanics to describe the behavior of particles at finite temperatures. In our metaphorical analysis, these functions could represent how the "temperature" (public interest or concern) affects the "particle behavior" (public opinion or information spread).

**Stable Green's Functions**: Would suggest that public interest or concern responds predictably to new information, which is desirable for stable information dissemination during disasters.

**Fluctuating Green's Functions**: Could indicate that public reaction is highly sensitive to changes in information, leading to unpredictable swings in public opinion or behavior.

In a "City" scenario, effective media strategies would aim for high FP Ghost counts (indicative of widespread engagement), narrow distributions of Indeterminate Ghosts (suggesting consistent information processing), low UV Divergence (indicating stable shifts in public opinion), and stable Green's functions (implying predictable public response to information). This would be indicative of an effective communication strategy during disasters, minimizing the impact of filter bubbles and ensuring that critical information reaches and is understood by the public.

## 10. Analysis of Results in a "Village" Scenario

The results appear to include a series of plots that can be interpreted as the evolution of opinions within a "Village" scenario, along with histograms for UV Divergence and Indeterminate Ghosts distribution. These could be used to analyze the dynamics of information spread within small, possibly less connected communities, which is critical when considering the dissemination of disaster and risk information through media.

### 10.1 Opinions Evolution

The opinions evolution plots typically show how individual opinions change over time. In a village scenario, where the population density is lower and the spread of information might be slower or less varied, the opinions might change less dramatically compared to a city scenario. The level of change would be indicative of the robustness of communication networks and the effectiveness of media in disseminating information.

### 10.2 UV Divergence Distribution

The UV Divergence histograms might represent the spread or variance in the change of opinions or information. In the context of media reporting:

**High UV Divergence**: Could suggest that certain news items or pieces of information have a significant impact on public opinion, potentially leading to rapid shifts that could be indicative of viral news or critical updates during a disaster.

**Low UV Divergence**: Might indicate a more uniform or controlled spread of information, which could be desirable in maintaining public order during emergencies.

### 10.3 Indeterminate Ghosts Distribution

The Indeterminate Ghosts histograms might represent the uncertainty or inconsistency in information spread.

**Wide Distribution**: Indicates a high variance in how information is received and processed by individuals, which in the context of a village might suggest a range of different interpretations or reactions to media reports, potentially due to a diversity of sources or the presence of misinformation.

**Narrow Distribution**: Suggests a more consistent and possibly accurate dissemination of information, which is essential for effective response during disasters.

### 10.4 Matsubara Green's Functions Perspective

While Matsubara Green's functions are specific to quantum statistical mechanics, if we draw a parallel to the dynamics of social systems, they could represent the response of a community to external "perturbations" such as news reports or disaster warnings.

**Stable Matsubara Green's Functions**: Would imply a consistent response to new information, with the community effectively integrating and acting on the information received.

**Fluctuating Matsubara Green's Functions**: Could suggest a community that is highly reactive to new information, which may lead to rapid but potentially uncoordinated responses to disasters.

In a "Village" scenario, it is crucial that disaster and risk information is conveyed effectively to ensure that all members of the community receive accurate and actionable information. The distribution of UV Divergence and Indeterminate

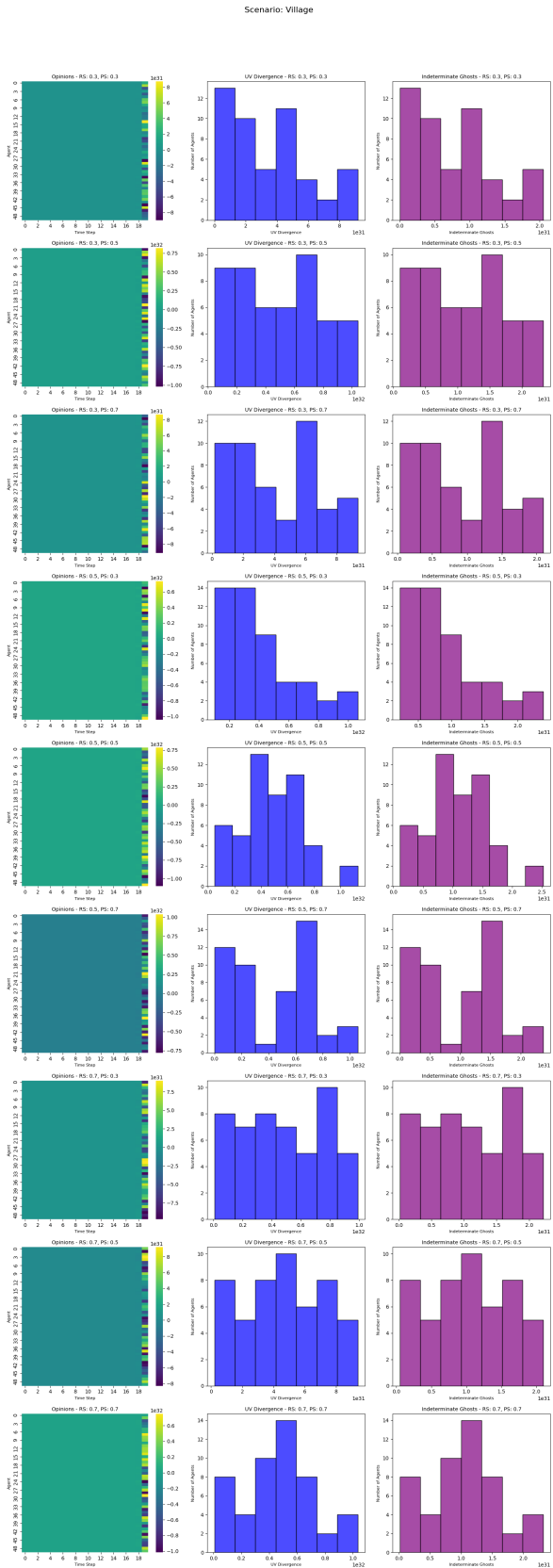

Fig. 16: Village:UV divergence, FP Ghost Distribution / Opinions Over Time

Ghosts provides insight into the spread and reception of information. Ideally, we would want to see low UV Divergence and narrow Indeterminate Ghost distributions, indicating a stable and uniform understanding of risk information. This would suggest that the media reporting strategy is effective, minimizing the impact of filter bubbles and ensuring community-wide comprehension and response to disaster-related communications.

# 11. Interpretation of Filter Bubbles in Observing Quantum Skipping Effects within Quantum Hall Effect

Finally, let's theoretically supplement the interpretation of filter bubbles and the concepts of annihilation and creation operators when observing quantum skipping effects within the context of the quantum Hall effect. We will explain this based on a general theoretical approach.

## 11.1 Interpretation of Filter Bubbles

In the framework of quantum field theory and relativistic quantum mechanics, the term *filter bubble* refers to regions where particles or exotic states are generated or annihilated. This concept is used to represent regions where the expected values of different fields in quantum field theory differ within the same space.

Within the context of filter bubbles, the *quantum skipping effect* refers to a phenomenon where quantum processes occur very rapidly within specific regions of a particular field. This suggests that the expected values or correlations of the field change rapidly within that region. As a result, different physical behaviors occur between regions where quantum effects are locally very strong and other regions.

## 11.2 Annihilation and Creation Operators

Annihilation operator $a_i$ and creation operator $a_i^\dagger$ are mathematical tools used to represent field operators in quantum field theory. These operators are used to describe the commutation relations of the Hamiltonian in quantum mechanics and field operators. Generally, creation operators represent the creation of new particles, while annihilation operators represent the annihilation of existing particles.

Specifically, in quantum field theory, you may consider a field operator $\phi(x)$, which represents the value of the field at position $x$. When expanding this field, creation and annihilation operators associated with different modes (such as wave numbers or energies) are introduced.

For example, in the field expansion of the quantum field $\phi(x)$, creation operator $a_k^\dagger$ and annihilation operator $a_k$ for mode $k$ can be expressed as follows:

$$\phi(x) = \int \frac{d^3k}{(2\pi)^{3/2}} \left( a_k e^{ikx} + a_k^\dagger e^{-ikx} \right)$$

Here, $a_k$ represents the annihilation of a particle in mode $k$, while $a_k^\dagger$ represents the creation of a particle in mode $k$.

## 11.3 Differences between Kubo's Formula and Matsubara Formalism

Kubo's formula is used to describe the response of a system to perturbations from external sources and is based on linear response theory. On the other hand, Matsubara formalism comprehensively describes the dynamics of a system using time-ordered Green's functions. The differences between these approaches are as follows:

**Kubo's Formula** It treats the impact of small perturbations from external sources on the system in a linear manner. This approach is used to understand how external information or stimuli in opinion formation among agents is processed.

**Matsubara Formalism** It employs time-ordered Green's functions to provide a comprehensive description of the entire dynamics of the system. This is a suitable method for analyzing the temporal progression of opinion exchange between agents in detail and its effects, considering nonlinear and quantum effects.

### Quantum Aharonov-Bohm effect(AB Effect), Potential Barriers (Coulomb Blockade), and Filter Bubbles

When the quantum Aharonov-Bohm effect(AB Effect) and potential barriers (Coulomb blockades) occur in the public discourse spaces of major cities or village communities, we will theoretically supplement and explain the interpretation of quantum field quantities for distant and proximity interactions, as well as the concepts of annihilation and creation operators. However, since detailed information about the quantum AB effect is not provided, we will explain it based on the framework of general quantum field theory.

### Quantum Aharonov-Bohm effect(AB Effect) and Potential Barriers (Coulomb Blockade)

The quantum Aharonov-Bohm effect(AB Effect) refers to the phase effects that occur in quantum mechanics when quantum particles take closed paths. On the other hand, potential barriers (Coulomb blockades) refer to the phenomenon where particles are unable to move due to the presence of an electric field. How these phenomena manifest in the public discourse spaces of major cities or village communities can be considered through the following theoretical approach.

## Distant and Proximity Interactions

In the public discourse spaces of major cities or village communities, interactions between agents can be both distant and proximate. Distant interactions refer to interactions that allow the transmission of information or opinions regardless of physical distance. In contrast, proximity interactions depend on physical closeness and take the form of dialogues or face-to-face communication.

When representing these interactions using quantum field theory, consider the quantum field $\phi(x)$. Distant and proximity interactions may have different physical effects, depending on the scenario.

## Annihilation and Creation Operators

Annihilation operator $a_i$ and creation operator $a_i^\dagger$ are mathematical tools used to represent field operators in quantum field theory. These operators are used to describe the commutation relations of field operators in quantum mechanics and represent the generation and annihilation of particles.

When considering the quantization of opinions or information exchange between agents, creation operator $a_i^\dagger$ represents the creation of new information or opinions, while annihilation operator $a_i$ represents the reduction or disappearance of information or opinions. The action of these operators depends on the specific states of agents or information.

## Theoretical Perspectives for Understanding Public Discourse Spaces

To understand the interactions and opinion formation among agents in public discourse spaces of major cities or village communities theoretically, the following perspectives can be considered:

(1) **Field Quantization**: Represent the information or opinions within the discourse space as a quantum field, using annihilation and creation operators to describe their changes.

(2) **Distant and Proximity Interactions**: Consider the differences between distant and proximity interactions, introducing interaction terms that depend on physical distance into the field theory.

(3) **AB Effect and Potential Barriers**: Investigate how the quantum AB effect and potential barriers analogous to Coulomb blockades affect the phase effects and information flow within the discourse space.

(4) **Description of Temporal Progression**: Analyze the temporal progression of opinion exchange among agents using Matsubara formalism or Green's functions to understand causality and correlations.

By combining these theoretical approaches, deeper insights into the opinion formation and interactions among agents in public discourse spaces of major cities or village communities can be gained. This may offer new perspectives for understanding social phenomena, making predictions, and formulating policies.

The explanations related to FP ghosts and ultraviolet divergences are provided as analogies within the context of public discourse spaces and information management. These analogies aim to relate concepts from quantum field theory to social communication and information exchange. While the specific details of quantum AB effects, Coulomb blockades, and these analogies may vary, they serve as imaginative tools for understanding the complex dynamics of information flow and communication in social systems.

## Coulomb Blockade and Filter Bubbles
### Coulomb Blockade

Coulomb blockade refers to a phenomenon in physics where charged particles are obstructed by potential energy barriers and remain unaffected.

### Filter Bubbles

Filter bubbles, on the other hand, denote areas of filtered information provided to individual users through information filtering algorithms or algorithms, creating a bubble of filtered information tailored to each user.

By connecting these two concepts, a pseudo-discussion can be developed as follows:

### Social Coulomb Blockade

Similar phenomena to Coulomb blockade can occur in social communication and information transmission. This refers to the state of being enclosed within information bubbles provided by specific information filters, algorithms, or platforms. In essence, individual users may access specific information sources or opinions while avoiding exposure to other information or opinions.

### Isolation of Discourse

This social Coulomb blockade can restrict the exchange of opinions and information between different information bubbles, potentially leading to a lack of diversity in opinions and multiple perspectives on information. This is analogous to communication being cut off from other information sources within filter bubbles. As a result, individuals exchange information and opinions within their own information ecosystems, with limited opportunities to access external information.

## Impact of Filter Bubbles

When filter bubbles cause a social Coulomb blockade, biases in information and opinions can emerge, leading to polarization of opinions and an imbalance of information. The constraint on diversity of discourse and the spread of information can result in distortions in social discussions and decision-making processes.

To address such phenomena, it is essential to break filter bubbles and ensure diversity of information, as well as facilitate the exchange of opinions. This could lead to a more equitable and diverse formation of opinions and information dissemination, potentially mitigating the effects of social Coulomb blockade.

## Coulomb Islands in the Discourse Space

The discussion of Coulomb islands in the context of discourse within filter bubbles is explained pseudo-theoretically. This discussion is based on the phenomenon of information filtering known as filter bubbles and serves to theoretically supplement the constraints on information and isolation of opinions within the discourse space.

## Filter Bubbles and Isolation of Discourse

Filter bubbles refer to the phenomenon where information tailored to individual users is provided through information filtering algorithms or platform algorithms. This information filtering can lead users to be exposed mainly to specific information or opinions while having limited access to others.

## Coulomb Islands in the Discourse Space

Coulomb islands within the discourse space refer to the phenomenon where individuals adhere to specific opinions or information and avoid contact with other information sources or differing opinions. This is related to the isolation of discourse caused by filter bubbles.

## Impacts of Coulomb Islands

The emergence of Coulomb islands within the discourse space can have the following impacts:

1. **Bias in Opinions**: Users who continuously engage with specific information sources or opinions are more likely to become biased towards that information or those opinions. This can result in a reduction of diversity in opinions, causing the discourse space to evolve in a biased direction.

2. **Distortion of Information**: Users trapped within filter bubbles are continuously exposed to specific information, leading to tailored advertisements and content. This can result in the distortion of users' awareness and information, making it difficult for them to access objective information and multiple perspectives.

3. **Isolation of Discourse**: Users trapped within Coulomb islands tend to distance themselves from dialogue or debates with differing opinions. This signifies constraints on opinion exchange and the closure of the discourse space.

## Countermeasures and Improvements

To mitigate the impact of Coulomb islands within the discourse space, the following measures can be considered:

1. **Ensuring Diversity of Information**: Filtering algorithms and platforms should provide users with opportunities to access diverse information and opinions.

2. **Transparency and Accountability**: Ensuring transparency and accountability in the operation of algorithms, allowing users to understand and control the information filtering process.

3. **Facilitating Opinion Exchange**: Encouraging opinion exchange and debates within the discourse space, providing a space for different opinions to interact, is essential.

By implementing these measures, the impact of social Coulomb blockade within the discourse space can be reduced, and a more equitable and diverse discourse environment can be established.

# 12. Conclusion: Media Reporting Patterns in Urban and Rural Environments

This section provides a detailed explanation of media reporting patterns in both large cities and rural communities. It also explores the phenomena associated with these patterns, including FP (False Positive) ghosts, Indeterminate Ghosts, and UV Divergence, considering them from the perspectives of Causal Green Functions, Delayed Green Functions, and Advanced Green Functions.

## 12.1 Causal Green Functions and Media Reporting

Causal Green Functions describe how past opinions influence the present. When applied to media reporting in different settings, such as large cities and rural areas, distinct scenarios emerge:

Large Cities: Rapid changes in information lead to strong short-term impacts reflected in Causal Green Functions. Past opinions are quickly replaced by new information. Rural Communities: Slower information dissemination results in Causal Green Functions showing that past opinions may have long-lasting effects on present opinion formation.

## 12.2 Delayed Green Functions and Information Reception

Delayed Green Functions indicate that past opinions influence present opinions with a time delay. This relates to how information is received:

Large Cities: Swift responses to new information often result in Delayed Green Functions showing short-term opinion changes. Rural Communities: Delayed Green Functions suggest that past opinions continue to influence the present due to slow information updates.

## 12.3 Advanced Green Functions and Information Prediction

Advanced Green Functions are employed for predicting future opinion changes, considering trends and forecasts regarding future events:

Large Cities: Frequent changes in trends and events place an emphasis on short-term predictions using Advanced Green Functions. Rural Communities: Stable information sources and environments make Advanced Green Functions valuable for long-term predictions.

## 12.4 Information Cutoff and Its Relation to Green Functions

Information cutoff involves filtering out misinformation or biased information and is closely related to Green Functions:

Large Cities: Swift information cutoff and filtering are necessary to accommodate short-term opinion changes. This aligns with the short-term movements indicated by Delayed Green Functions and Advanced Green Functions. Rural Communities: Careful information cutoff is essential to maintain long-term opinion consistency. As suggested by Causal Green Functions, past opinions may have a strong impact on the present, necessitating a cautious approach to information updates.

In summary, urban and rural areas require different approaches to address the influence of media, information reception, and information cutoff. Utilizing Green Functions helps us better understand these differences, enabling the development of appropriate information filtering strategies.

## 13. Conclusion: Roles of Mass Media and Social Media in Disaster Reporting

Here, we would like to discuss media coverage of disasters, including major earthquakes and other emergencies, as well as the increasing number of cases of disaster victims themselves being reported by the media in recent years.

In this section, we will discuss the roles of mass media and social media in disaster reporting, considering the occurrence of False Positive (FP) Ghosts, Indeterminate Ghosts, and UV Divergence in both urban and rural environments.

We will also provide theoretical insights from the perspectives of Causal Green Functions, Delayed Green Functions, and Advanced Green Functions.

## 13.1 Roles of Mass Media and Social Media

**Mass Media**: Mass media is generally considered an official source of information and plays a role in providing reliable information during disasters. **Social Media (SNS)**: Social media allows real-time information sharing and can provide direct reports from eyewitnesses and victims during disasters.

## 13.2 Perspective of Causal Green Functions

Causal Green Functions illustrate how past information influences present opinions. In the context of disaster reporting:

**Urban Environments**: With less past experience of disasters, there may be stronger reactions to new reporting. **Rural Communities**: Past disaster experiences may lead to a more cautious reception of information.

## 13.3 Perspective of Delayed Green Functions

Delayed Green Functions show how much current opinions are influenced by past information. In disaster reporting, initial reporting can have a significant impact on subsequent opinion formation:

**Urban Environments**: Early reporting can potentially induce opinion changes. **Rural Communities**: Consistent reactions to early reporting may be observed.

## 13.4 Perspective of Advanced Green Functions

Advanced Green Functions are used to predict future opinion changes. In disaster reporting, the influence of early information on future opinion formation is crucial:

**Urban Environments**: Quick incorporation of new information may lead to rapid opinion changes. **Rural Communities**: Evaluating information with long-term effects in mind is common.

## 13.5 FP Ghosts, Indeterminate Ghosts, and UV Divergence

**FP Ghosts**: The interaction with SNS and mass media intensifies during disasters, leading to an increase in FP Ghosts. **Indeterminate Ghosts**: In cases of conflicting or uncertain information, the consistency of opinions may decrease, resulting in an increase in Indeterminate Ghosts. **UV Divergence**: Rapid changes in new information or situations may result in significant opinion fluctuations.

Differences in response to disaster reporting between urban and rural areas can be understood through the lenses of Causal Green Functions, Delayed Green Functions, and Advanced Green Functions. FP Ghosts, Indeterminate Ghosts, and UV Divergence serve as important indicators of the dynamics of these interactions. Understanding variances in information reception and response in disaster reporting enables the formulation of more effective communication strategies.

# 14. Conclusion: Roles of Mass Media and Social Media in Disaster Reporting

When considering the reporting of volunteer requests and support information during major disasters, the roles of mass media and social media differ, and each conveys information in different patterns in urban and rural communities. In conjunction with this, we will examine the occurrence of FP Ghosts, Indeterminate Ghosts, and UV Divergence from the perspectives of Causal Green Functions, Delayed Green Functions, and Advanced Green Functions.

## 14.1 Roles of Mass Media and SNS

**Mass Media**: Generally provides official information and conveys volunteer requests and support information officially. Especially in large cities, it can reach a wide audience. **Social Media (SNS)**: Allows individual users to share information in real-time and quickly spread emergency needs and support requests. Particularly effective for sharing information within local communities in rural areas.

## 14.2 Perspective of Causal Green Functions

Causal Green Functions illustrate how past information influences reactions to the current situation.

**Urban Environments**: Past experiences of disasters influence current reactions. For example, strong responses may occur towards support methods that were effective in the past. **Rural Communities**: Past community experiences influence current support activities, with an emphasis on traditional support methods and organizations.

## 14.3 Perspective of Delayed Green Functions

Delayed Green Functions show how much current information is influenced by past reactions.

**Urban Environments**: Quick responses to new information are important, and current information is prioritized over past experiences. **Rural Communities**: Information is accepted based on past experiences, and responses to new information tend to be cautious.

## 14.4 Perspective of Advanced Green Functions

Advanced Green Functions are used to predict changes in future information. **Urban Environments**: Rapid predictions and responses based on current information are made for future support needs and disaster response plans. **Rural Communities**: Long-term planning for future support and disaster preparedness is emphasized, often based on regional traditions and experiences.

## 14.5 Perspective on FP Ghosts, Indeterminate Ghosts, and UV Divergence

- **FP Ghosts**: Frequent interaction with mass media and SNS during disasters leads to the occurrence of FP Ghosts. In large cities, information from various media sources increases, while in rural communities, information is limited. **Indeterminate Ghosts**: In cases of conflicting or uncertain reporting, the consistency of opinion changes decreases, especially leading to an increase in Indeterminate Ghosts in large cities. **UV Divergence**: Rapid reactions and information changes in disaster reporting lead to UV Divergence. In large cities, the speed and diversity of information distribution make UV Divergence more likely.

In large cities and rural communities, responses to disaster reporting differ, and mass media and SNS play different roles. FP Ghosts, Indeterminate Ghosts, and UV Divergence serve as essential indicators of the dynamics of these responses. Understanding the differences in information reception and reactions in disaster reporting enables the formulation of more effective communication strategies.

# 15. Roles of Remote and Proximate Interactions in Disaster Reporting

In the context of conveying volunteer requests and support information during major disasters, both remote and proximate interactions play crucial roles. To better understand the dynamics of these interactions, we will delve further into the discussion from the perspectives of Causal Green Functions, Delayed Green Functions, and Advanced Green Functions.

## 15.1 Perspective of Remote Interactions

Remote interactions refer to the process of information transmission regardless of physical distance, which is particularly vital in media reporting during major disasters.

**Mass Media**: Mass media such as television, radio, online news, etc., provide information through extensive remote interactions. This enables large-scale volunteer mobilization and distribution of support materials. **Social Media (SNS)**: SNS, where individuals share real-time information, is a powerful tool for remote interactions. People located far from disaster-stricken areas can participate in support activities and spread information through SNS.

## 15.2 Perspective of Proximate Interactions

Proximate interactions involve information transmission based on physical proximity, emphasizing the importance of local communities and direct human relationships. **Local Communities**: Communities in disaster-affected areas strengthen proximate interactions through face-to-face information sharing and support activities. This results in locally tailored and specific support. **Individual Networks**: Personal social networks also play a significant role in proximate interactions. Information sharing among friends, family, and neighbors contributes to identifying support needs and resource allocation.

### 15.3 Comprehensive Perspective

**Causal Green Functions**: Past disaster experiences and existing media usage patterns influence current responses. Both remote and proximate interactions play a role. **Delayed Green Functions**: Demonstrates how existing information networks are utilized in information collection and responses immediately following a disaster. The rapid transmission of information through remote interactions and the local community's response through proximate interactions are crucial. **Advanced Green Functions**: The role of information in future support activities and disaster response plans, with particular significance from the influence of information flow through remote interactions.

In the transmission of volunteer requests and support information during major disasters, mass media and SNS play crucial roles through remote interactions, while local communities and personal networks enhance proximate interactions. Understanding these dynamics is essential for effective disaster response and support planning. By analyzing the dynamics of these interactions using Causal Green Functions, Delayed Green Functions, and Advanced Green Functions, a more comprehensive disaster response strategy can be formulated.

Here, we will discuss extreme cases of media coverage of disasters, such as major earthquakes and other emergencies, and the increasing number of cases of disaster victims themselves being reported by the media.

### 15.4 Quantum AB Effect (AharonovBohm Effect)

The quantum AB effect is a phenomenon where changes in the electromagnetic potential surrounding a region, even in the absence of an electromagnetic field, influence quantum systems. In a social context, this effect can be metaphorically used to describe the "invisible influence" – how media's impact and shifts in public opinion, which are not directly observed, indirectly affect societal opinion formation and behavior.

### 15.5 Hall Effect

The Hall effect is a phenomenon where applying a magnetic field to a conductor carrying an electric current results in the generation of voltage within the conductor. In a social context, external influences (e.g., information from mass media or social media) affecting a social group and causing unexpected reactions or changes (analogous to "social voltage") can be likened to the Hall effect.

### 15.6 Coulomb Blockade

Coulomb blockade is a phenomenon that occurs in very small conductors (e.g., quantum dots) where charges obstruct the flow of other charges. In a social context, this can serve as a metaphor for describing situations where specific information or misinformation inhibits the flow of communication within a social group. For instance, it suggests that biased reporting may impede the circulation of alternative viewpoints or information.

Hypothetically applying the concepts of the quantum AB effect, Hall effect, and Coulomb blockade from physics to the roles of mass media and social media in disaster reporting provides a metaphorical interpretation. Using these concepts, let's theoretically supplement the understanding of media influence in urban and rural communities.

### 15.7 Quantum AB Effect and Media Reporting

The quantum AB effect demonstrates how an invisible electromagnetic field affects the behavior of matter. When applied to media reporting, it suggests the existence of media with indirect yet substantial influence.

Urban Areas: Diverse media sources exist, indirectly influencing public opinion and behavior. High information diversity can result in indirect influences similar to the AB effect. Rural Communities: Limited information sources may have strong indirect influences on the community. This is akin to the AB effect, where specific information has a more direct and noticeable impact.

### 15.8 Hall Effect and Information Flow

The Hall effect shows how external magnetic fields influence charges within a material. In the context of media reporting, it can metaphorically represent how external information influences societal opinions and behaviors.

Urban Areas: Various information sources affect societal opinions and actions, resembling the Hall effect with diverse external influences generating varied responses. Rural Communities: Due to limited information sources, specific media outlets may exert strong influence, similar to the Hall effect where specific influences are prominent.

### 15.9 Coulomb Blockade and Information Circulation

Coulomb blockade describes how charges in small conductors obstruct the flow of other charges. In the context of media reporting, this can symbolize scenarios where specific information hinders the flow of communication within a society.

Urban Areas: Many information sources exist, but specific messages or misinformation can block the circulation of other information. This is akin to Coulomb blockade, where certain information restricts information flow. Rural Communities: Limited information sources make it easier for specific information to dominate communication, resembling Coulomb blockade where information flow can be constrained.

Applying these physics concepts hypothetically to media analysis offers a new perspective on understanding the dynamics of social communication. However, it's essential to note that these are metaphorical approaches, and rigorous research based on specific data and theories is required in actual social science analysis.

# Aknowlegement

The author is grateful for discussion with Prof. Serge Galam and Prof.Akira Ishii.